\documentclass[12pt]{article}
\usepackage{}
\usepackage{geometry}             
\usepackage{graphicx}
\usepackage{amssymb}
\usepackage{amsmath}
\usepackage{epstopdf}
\usepackage{comment}
\usepackage{cite}
\usepackage{caption}

\usepackage[hyperindex=true,
          pdfstartview=FitH,
          bookmarksnumbered=true,
          bookmarksopen=true,
          citecolor=blue,
          linkcolor=blue,
          colorlinks=true,
          pdfborder=001,
          unicode]{hyperref}

\allowdisplaybreaks

\begin{document}
\title{Gauss-Bonnet coupling constant
as a free thermodynamical variable
\\and the associated criticality}
\author{Wei Xu, Hao Xu and Liu Zhao\\
School of Physics, Nankai University, \\
Tianjin 300071, P R China\\
{\em email}: \href{mailto:xuweifuture@mail.nankai.edu.cn}{xuweifuture@mail.nankai.edu.cn}, \\
\href{mailto:physicshx@gmail.com}{physicshx@gmail.com} and \href{mailto:lzhao@nankai.edu.cn}{lzhao@nankai.edu.cn}}
\date{}
\maketitle

\begin{abstract}
The thermodynamic phase space of Gauss-Bonnet (GB) AdS black holes is extended,
taking the inverse of the GB coupling constant as a new thermodynamic
pressure $P_{\mathrm{GB}}$. We studied the critical behavior associated with
$P_{\mathrm{GB}}$ in the extended thermodynamic phase space at fixed
cosmological constant and electric charge. The result shows that when the black
holes are neutral, the associated critical points can only exist in five
dimensional GB-AdS black holes with spherical topology, and the corresponding
critical exponents are identical to those for Van der Waals system. For charged
GB-AdS black holes, it is shown that there can be only one critical point
in five dimensions (for black holes with either spherical or hyperbolic
topologies), which also requires the electric charge to be bounded
within some appropriate range; while in $d>5$ dimensions, there can be up to
two different critical points at the same electric charge, and the phase
transition can occur only at temperatures which are not in between the two
critical values.
\end{abstract}

\section{Introduction}

Thermodynamic properties of black holes have been studied for many years,
especially in anti-de Sitter (AdS) spacetime due to the AdS/CFT correspondence
\cite{Maldacena:1997re,Gubser:1998bc,Witten:1998qj,Brown:1986nw}. An
outstanding feature for AdS black holes is the so-called Hawking-Page phase
transition, which can happen between stable large black holes and thermal gas
in AdS spacetime \cite{Hawking:1982dh}. Thermodynamics of charged black holes
in AdS spacetime has also been intensively studied. The asymptotically AdS
charged black holes admit a gauge duality description via a dual thermal field
theory. This dual description suggests that charged AdS black holes
exhibit critical behavior in the $Q-\phi$ diagram which is reminiscent to the
liquid-gas phase transitions in a Van der Waals system
\cite{Chamblin:1999tk,Chamblin:1999hg,Tsai:2011gv,Niu:2011tb,
Banerjee:2011cz,Banerjee:2012zm,Lala:2012jp,Wei:2012ui}, where $Q$ and $\phi$
denote the electric charge and potential, respectively.

Recently, this picture has been substantially extended. The idea of including
the variation of the cosmological constant $\Lambda$ in the first law of black
hole thermodynamics has acquired increasing attention
\cite{Caldarelli:1999xj,Kastor:2009wy,Dolan:2010ha,Dolan:2011xt,
Dolan:2011jm,Dolan:2013dga,Cvetic:2010jb,Lu:2012xu}. By studying critical
behaviors of AdS black holes in the extended phase space, i.e. the $P-V$
diagram \cite{Kubiznak:2012wp,Gunasekaran:2012dq,
Dolan:2012jh,Belhaj:2012bg,Hendi:2012um,Chen:2013ce,Hristov:2013sya,
Spallucci:2013osa,Zhao:2013oza,Belhaj:2013ioa,Poshteh:2013pba,
Cai:2013qga,Belhaj:2013cva,Altamirano:2013uqa}, where $P$ is the thermodynamic
pressure associated with the
cosmological constant, which takes the value
\begin{align}
    P=-\frac1 {8\pi}\Lambda=\frac{(d-1)(d-2)}{16\pi \ell^2}
\label{pressure1}
\end{align}
in the geometric units $G_N =\hbar=c=k=1$, with $\ell$ being the
$d$-dimensional AdS radius, $V$ is the conjugate ``thermodynamic volume"
\cite{Dolan:2011xt,Cvetic:2010jb,Dolan:2013ft,Ballik:2013uia}, the
analogy in the
$Q-\phi$ diagram of AdS charged black hole as a Van der Waals system has
been further enhanced. Both systems share the same critical exponents and have
extremely similar phase diagrams. This analogy has been generalized to the
higher dimensional charged black holes \cite{Kubiznak:2012wp,
Belhaj:2012bg,Spallucci:2013osa}, rotating black holes
\cite{Poshteh:2013pba,Belhaj:2013cva,Altamirano:2013uqa},
Gauss-Bonnet (GB) black
holes \cite{Wei:2012ui,Cai:2013qga}, $f(R)$ black hole\cite{Chen:2013ce}, black
holes with scalar hair\cite{Hristov:2013sya,Belhaj:2013ioa}, black holes with
nonlinear source\cite{Hendi:2012um}, and Born-Infeld black holes
\cite{Gunasekaran:2012dq} in AdS space.

The inclusion of $P-V$ variables in the thermodynamics of AdS black holes
is not just an artificial game to play with. There exist several reasons why
$\Lambda$ should be included as a thermodynamical variable
\cite{Kubiznak:2012wp}. One may suppose
that there are some more fundamental theory in which some physical constants
such as Yukawa and gauge coupling constants, Newton's constant, or
cosmological constant, may not be fixed values but dynamical ones arising from
the vacuum expectation values and hence can vary. Thus, it is natural to add
variations of these ``constants'' into the first law of black hole
thermodynamics \cite{Gibbons:1996af,Creighton:1995au}. Besides, the
cosmological constant term is necessary in the first law of black hole
thermodynamics, in order to get a consistent Smarr relation for balck hole
thermodynamics from the scaling argument \cite{Kastor:2009wy}. Similar
situation appears for the Born-Infeld black holes \cite{Rasheed:1997ns,
Breton:2004qa,Huan:2010} and GB black holes \cite{Cai:2013qga}. To
get a consistent Smarr relation by scaling arguments, one has to introduce the
Born-Infeld and GB parameter terms into the first law of black holes.
Once one takes the cosmological constant as thermodynamic pressure in the
first law, the black hole mass $M$ should be explained as enthalpy rather than
internal energy of the system \cite{Kastor:2009wy}.

In this paper, we will revisit the first law of thermodynamics for GB
black holes with emphasis on the role of GB coupling $\alpha$. It is
well known that GB gravity theory has two AdS solutions with
effective cosmological constants \cite{Boulware:1985wk,Cai:2001dz,
Dehghani:2004vn,Xu:2011sja,Zhao:2011px}
\begin{align}
    \Lambda_{\mathrm{eff}}=-\frac{\, \left( d-1 \right)  \left( d-2 \right)}
    {4\tilde{\alpha}}  \left( 1-\delta\,\sqrt
{1+8\,{\frac {\tilde{\alpha}\,\Lambda}{ \left( d-1 \right)  \left( d-2
 \right) }}} \right), \label{leff}
\end{align}
where $\tilde{\alpha}=(d-3)(d-4)\alpha$. $\delta=+1$ corresponds to the general
relativistic (GR) branch and has a GR limit as $\alpha\rightarrow0$,
while $\delta=-1$ corresponds
to the GB branch and does not have the GR limit. It has been shown by
Boulware and Deser that the GB branch is unstable and the graviton is
a ghost, while the GR branch is stable and is free of ghosts
\cite{Boulware:1985wk}. Thus we shall only consider the GR branch with
$\delta=+1$. By a simple analogy with the previously mentioned $P-V$
criticality for AdS black holes, one may tend to consider
$P_{\mathrm{eff}}\propto -\Lambda_{\mathrm{eff}}$ as a thermodynamic pressure.
However, for two reasons, we will not take this point of view. Firstly,
$\Lambda_{\mathrm{eff}}$ is a complicated combination of {\em two}
parameters $\Lambda$ and $\alpha$ and we wish to understand the role of each
parameter {\em independently}. Secondly, in the enthalpy description of the
first law, $P_{\mathrm{eff}}$ always appear in differential form, i.e. in the
term $V_{\mathrm{eff}} \,\mathrm{d}P_{\mathrm{eff}}$, and it is reasonable to
decompose $\mathrm{d}P_{\mathrm{eff}}$ as a combination of $\mathrm{d}\Lambda$
and $\mathrm{d}\alpha$, thanks to from (\ref{leff}). Moreover, we have a
good reason to consider
\begin{align}
P_{\mathrm{GB}}\equiv \frac{1}{8\pi\alpha} \label{pgb}
\end{align}
instead of $\alpha$ as a component of thermodynamic pressure, because it is
$\frac{1}{\alpha}$ which scales like a pressure. This is where our
considerations depart from the previous work \cite{Cai:2013qga}, which took
$\alpha$ and its conjugate $\mathcal{A}$ as a pair of thermodynamic quantities. 
This consideration is also different from the geometric formulation for 
Lovelock theories \cite{Kastor:2010gq} which contain Gauss-Bonnet gravity as a 
special case. Nonetheless, it will become clear that using $P_{\mathrm{GB}}$ 
and its conjugate as a pair of thermodynamic variables indeed reveals novel 
criticalties of the theory which is otherwise difficult to describe in 
terms of $\alpha$-$\mathcal{A}$ variables.

We will consider the criticality associated with the new thermodynamic
variables $P_{\mathrm{GB}}$ and its conjugate $V_{\mathrm{GB}}$ in the
extended phase space of charged GB-AdS black holes.
It will be shown that this kind of critical behavior is
different from the previously mentioned $P-V$ criticality, and that the spacial
curvature of the AdS black hole horizon plays an important role in such
criticalities.

The paper is organized as follows. Section 2 is devoted to the thermodynamics
of GB-AdS black holes in the extended phase space with
$P_{\mathrm{GB}}$ and $V_{\mathrm{GB}}$. In section 3, we consider the
criticality associated with $P_{\mathrm{GB}}$ for static neutral GB-AdS
black holes. The case for static charged GB-AdS black holes is discussed in
section 4. Finally, some concluding remarks are given in the last section.

\section{Extended thermodynamics of GB-AdS black holes}

The action of $d$-dimensional Einstein-GB-Maxwell theory with a bare
cosmological constant $\Lambda$ reads
\begin{align}
I=\frac{1}{16\pi}\int \mathrm{d}^dx \sqrt{-g}\bigg[R-2\Lambda+\alpha (R_{\mu\nu
\gamma\delta}R^{\mu\nu\gamma\delta}-4R_{\mu\nu}R^{\mu\nu}+R^2)\bigg]
-\frac{1}{4}\int \mathrm{d}^dx \sqrt{-g} F_{\mu\nu}F^{\mu\nu},
\label{action}
\end{align}
where the GB coupling $\alpha$ has dimension $[{\rm length}]^2$ and
can be identified with the inverse string tension with positive value
\cite{Boulware:1985wk} if the theory is incorporated in string theory, thus
we shall consider only the case $\alpha>0$. Of course, we take the
spacetime dimension $d \ge 5$, since in $d=4$ dimensions, the integration of
the GB density $\mathcal{L}_{\rm GB}=R_{\mu\nu\gamma\delta}
R^{\mu\nu\gamma\delta}-4R_{\mu\nu}
R^{\mu\nu}+R^2$ is a topological number and has no dynamics.

The $d$-dimensional static charged GB-AdS black hole solution arising from the
field equations associated with the action (\ref{action}) is known to take the
form
\begin{align}
\mathrm{d}s^2=-f(r)\mathrm{d}t^2+\frac{1}{f(r)}\mathrm{d}r^2
+\mathrm{d}\Omega_{d-2,k}^2,
 \label{metric}
\end{align}
which should be accompanied by a standard Coulomb
potential for the Maxwell field, where $\mathrm{d}\Omega_{d-2,k}^2$ represents
the line element of a $(d-2)$-dimensional maximally symmetric Einstein space
with constant curvature $(d-2)(d-3)k$, where $k = 1$, $0$ and $-1$ correspond
to the spherical, Ricci flat
and hyperbolic topology of the black hole horizons, respectively. The metric
function $f(r)$ is given by \cite{Boulware:1985wk,Cai:2001dz,
Wiltshire:1985us,Cvetic:2001bk}
\begin{align}
f(r)=k+\frac{r^2}{2\tilde{\alpha}}\left (1-\sqrt{1-\frac{4\tilde{\alpha}}
{\ell^2}+\frac{64\pi\tilde{\alpha} M}{(d-2) r^{d-1}}-\frac{2\tilde{\alpha} Q^2}
{(d-2)(d-3)r^{2d-4}}} \right ),
\label{f(r)}
\end{align}
where $M$ and $Q$ are the mass and charge (in geometric unit) of the black 
hole, respectively.

Since we are going to discuss the thermodynamics of the black hole in the
extended phase space by introducing extra thermodynamical variables $P-V$ and
$P_{\mathrm{GB}}-V_{\mathrm{GB}}$, the black hole mass $M$ should be identified
with the enthalpy $H\equiv M$ rather than the internal energy of the
gravitational system \cite{Kastor:2009wy}. It follows from (\ref{f(r)}) that
$H$ can be expressed in terms of the horizon radius $r_+$
\begin{align}
H=\left[\frac{(d-2) r_+ ^{d-3}}{16\pi}\left (k+\frac{r_+^2}{\ell^2}+\frac{k^2}
{8\pi r_+^2P_{\mathrm{GB}}}\right )+\frac{Q^2}{32\pi (d-3)r_+^{d-3}}\right],
\label{enthalpy}
\end{align}
where $r_+$ is the largest root of $f(r)$. The Hawking temperature of the black
hole is given by
\begin{align}
T=\frac1{4\pi}f'(r_+)=\frac{\frac{(d-1)r_+^4}{\ell^2}+(d-3)k r_+ ^2+
\frac{(d-5)k^2}{8\pi P_{\mathrm{GB}}}-\frac{Q^2}{2(d-2)r_+ ^{2d-8}}}{4\pi r_+
(r_+ ^2+\frac{k}{4\pi P_{\mathrm{GB}}})}.
\label{temperature}
\end{align}
Other thermodynamic quantities are well known in the literature
\cite{Cai:2001dz,Cvetic:2001bk}. For examples, the entropy $S$ and electric
potential $\Phi$ are given by
\begin{align}
&S=\frac{ r^{d-2}_+}{4}\left (1+\frac{(d-2) k}{4\pi (d-4)r_+^2
P_{\mathrm{GB}}}\right ),
\label{entropy}\\
&\Phi=\frac{ Q}{8\pi (d-3)r_+^{d-3}}.
\label{phi}
\end{align}
Note that here we have treated the cosmological constant and GB
coupling constant as free thermodynamical variables, their conjugate quantity
thermodynamic volumes $V$ and $V_{\mathrm{GB}}$ are given respectively by
\begin{align}
&V=\left(\frac{\partial H}{\partial P}\right)_{S,Q,P_{\mathrm{GB}}}
=\frac{ r_+ ^{d-1}}{d-1},
\label{volume}\\
&V_{\mathrm{GB}} \equiv \left(\frac{\partial H}{\partial
P_{\mathrm{GB}}}\right)_{S,Q,P}=-\frac{(d-2)k^2}
{128\pi^2 P_{\mathrm{GB}}^2} r^{d-5}_+.
\label{VGB}
\end{align}
Note that these thermodynamic volumes are in no way linked to the geometric
volume of the black hole: in fact, from the point of view of static observers
located outside of the black hole, there is no such notion of a geometric
volume of the black hole at all. The physical meaning of these thermodynamic
volumes remains obscure. Therefore, the negative value of $V_{\mathrm{GB}}$
should not be considered unacceptable. Anyway, the negativity of
$V_{\mathrm{GB}}$ does not prevent us from studying
criticality in the extended thermodynamic phase space, since in the following
we shall take $r_+$ -- which is always positive -- instead
of $V_{\mathrm{GB}}$
as an equation of state (EOS) parameter associated with $P_{\mathrm{GB}}$.

An important relation in black hole thermodynamics is Smarr relation. In order
to obtain the correct Smarr relation for the
above-mentioned charged static GB-
AdS black hole, we will now make some scaling arguments. Since the black hole
enthalpy $H$ (i.e. the mass $M$) is a homogeneous function of entropy $S$,
electric charge $Q$ and thermodynamic pressures $P$ and $P_{\mathrm{GB}}$, and
that $M$ scales as $[\mathrm{length}]^{d-3}$, $S$ scales as
$[\mathrm{length}]^{d-2}$, $Q$ scales as $[\mathrm{length}]^{d-3}$, $P$ and
$P_{\mathrm{GB}}$ scales as $[\mathrm{length}]^{-2}$ (see, e.g.
\cite{Kastor:2009wy,Kastor:2010gq}), we find that the Smarr relation for the
black hole under consideration reads
\begin{align}
  (d-3)H=(d-2)TS+(d-3)Q\Phi-2PV-2P_{\mathrm{GB}}V_{\mathrm{GB}}.
\label{Smarr}
\end{align}
This is certainly different from the Smarr relation known from previous
literature, because we have now extended the thermodynamic phase space and
have taken $P_{\mathrm{GB}}$ and $V_{\mathrm{GB}}$ as a new pair of conjugate
thermodynamic variables. Correspondingly, the first law of black hole
thermodynamics is generalized as
\begin{align}
\mathrm{d}H= T\mathrm{d}S +\Phi \mathrm{d}Q +V\mathrm{d}P +V_{\mathrm{GB}}
\mathrm{d}P_{\mathrm{GB}}.\label{firstlaw}
\end{align}

In the rest of this paper, we shall consider criticality associated with
the new variable $P_{\mathrm{GB}}$, taking $P$ and $Q$ as fixed
parameters. In order to study the criticality of black holes, it is necessary
to work with the Gibbs free energy $G$, which can be obtained via the Legendre
transformation
\begin{align}
G&=G(T,Q,P,P_{\mathrm{GB}})=H-TS\nonumber\\
&=\frac{r_+^{d-3}(d-2)}{16\pi}\left(k+\frac{r_+^2}{\ell^2}\right)
-\frac{\,Tr_+^{d-2}}{4} \left( 1+\,{\frac { \left( d-2 \right) \,k}{4\pi
P_{\mathrm{GB}} \left( d-4 \right) r_+^{2}}} \right)\nonumber\\
&+\frac{k^2(d-2)r_+^{d-5}}{128\pi^2P_{\mathrm{GB}}}
+\,{\frac {{Q}^{2}r_+^{3-d}}{32\pi \, \left( d-3 \right) }}.
\label{Gibbs}
\end{align}
We shall also make use of the internal energy
\begin{align}
U&=H-P_{\mathrm{GB}}V_{\mathrm{GB}}\nonumber\\
&=\frac{(d-2) r_+ ^{d-3}}{16\pi}\left (k+\frac{r_+^2}{\ell^2}+\frac{k^2}{4\pi
r_+^2P_{\mathrm{GB}}}\right )+\frac{Q^2}{32\pi (d-3)r_+^{d-3}}
\label{IE}
\end{align}
while evaluating the heat capacity at fixed $r_+$, where $P_{\mathrm{GB}}$ is
to be taken as a function of the temperature $T$, thanks to the EOS
\begin{align}
 P_{\mathrm{GB}}=\frac{ \left((5-d)k+8\,T\pi \,r_+ \right) (d-2)k}{8\pi
 \left[\left( {\frac {r_+^2 \left( d-1 \right) }{{\ell}^{2}}}+(d-3)k-4\,T
\pi \,r_+ \right) (d-2)r_+^{2}-\frac{Q^2}{2r_+^{2d-8}}\right]}
\label{PGB2}
\end{align}
which arises from (\ref{temperature}).
Note that for $k=0$, $P_{\mathrm{GB}}$ is identically zero, thus
$P_{\mathrm{GB}}$ and $V_{\mathrm{GB}}$ loses their role as a pair of
thermodynamic variables. So, in this paper, we shall always assume $k\neq 0$.
This EOS is quite different from the Van der Waals equation
\begin{align}
    P=\frac T{v-b}-\frac a{v^2} \label{Van}
\end{align}
which states that the pressure $P$ is a linear function of the temperature $T$,
if the specific volume $v$ is kept fixed. Nonetheless, the usual method for
studying criticalities for Van der Waals system still works, as will be seen
below.

Before proceeding, let us note that there are some natural constraints on the
allowed range of $P_{\mathrm{GB}}$: Firstly, a well-defined vacuum solution
with $M=0,Q=0$ results in \cite{Cai:2013qga} $P_{\mathrm{GB}}\geq\frac{1}{2\pi
\ell^2}$ (i.e. the dimensionless pressure $\frac{P_{\mathrm{GB}}}{P}$ has a 
lower bound: $\frac{P_{\mathrm{GB}}}{P}\geq\frac{8}{(d-1)(d-2)}$); Secondly, 
the non-negative definiteness
of the black hole entropy (\ref{entropy}) gives another constraint
\cite{Clunan:2004tb} $P_{\mathrm{GB}}\geq\frac{d-2}{4(d-4)\pi r_+^2}$
for $k=-1$.

\section{Criticality for static neutral GB-AdS black holes associated with
$P_{\mathrm{GB}}$}

\subsection{Critical behavior in five and six dimensions}

For simplicity, we begin our study on the case of neutral black holes. The
black hole enthalpy (\ref{enthalpy}) is reduced to
\begin{align}
H=\frac{(d-2) r_+ ^{d-3}}{16\pi}\left (k+\frac{r_+^2}{\ell^2}+\frac{k^2}{8\pi
P_{\mathrm{GB}}r_+^2}\right ),
\end{align}
and the Gibbs free energy becomes
\begin{align}
G=\frac{(d-2) r_+ ^{d-3}}{16\pi}\left (k+\frac{r_+^2}{\ell^2}+\frac{k^2}{8\pi
P_{\mathrm{GB}}r_+^2}\right )-\frac{\,Tr_+^{d-2}}{4}
\left( 1+\,{\frac { \left( d-2 \right)\,k}{4\pi \left( d-4 \right) r_+^{2}
P_{\mathrm{GB}}}}
 \right).\label{Gibbs1}
\end{align}
The EOS of the black holes simplifies into
\begin{align}
P_{\mathrm{GB}}=\frac{\left((5-d)k+8\,T\pi \,r_+ \right) {k}}{8\pi\left( {\frac
{r_+^2 \left( d-1 \right) }{{\ell}^{2}}}+(d-3)k-4\,T
\pi \,r_+ \right) r_+^{2}}.
\label{PGB1}
\end{align}

The critical point is determined as the inflection point on the
$P_{\mathrm{GB}}-r_+$ diagram, i.e.,
\begin{align}
\left. \frac{\partial P_{\mathrm{GB}}}{\partial r_+}\right |_{r_+=r_{c},T=T_c}=
\left. \frac{\partial^2 P_{\mathrm{GB}}}{{\partial r_+^2}}
\right |_{r_+=r_{c},T=T_c}=0,
\label{critialpont}
\end{align}
and $\frac{\partial^2 P_{\mathrm{GB}}}{{\partial r_+^2}}
\big|_{r_+=r_{c}+0^+,T=T_c}$ and $\frac{\partial^2 P_{\mathrm{GB}}}
{{\partial r_+^2}}\big |_{r_+=r_{c}+0^-,T=T_c}$ should have different signs,
where we have used the subscript $c$ to stand for the quantities at the
critical point. Using (\ref{PGB1}), the two conditions in (\ref{critialpont})
becomes a pair of very complicated algebraic equations for $r_c$ and $T_c$,
which, of course, depend on the spacetime dimension $d$ and the signature of
the spacial curvature $k$ of the black hole horizons. For $k=-1$, the pair of
equations arising
from (\ref{critialpont}) can never have a solution with real and positive
$r_c$. Therefore, we are left with only the choice $k=+1$. In this case,
eliminating $T_c$ from the above pair of equations,
we get a single simplified equation determining the critical radius $r_c$,
\begin{align}
 36(d-1)^2R_c^2-12(d-1)(2d-9)R_c+(d-3)(7d-39)= 0,
\label{eqc}
\end{align}
where we have introduced
\begin{align}
  R_c=\frac{r_c^2}{\ell^2}.
\end{align}
The solutions to this equation read
\begin{align}
  R_c=\,{\frac { 2\,d-9\pm\,\sqrt {-3(d-2)(d-6)} }{6(d-1)}}. \label{RC}
\end{align}
In order to find a real positive $R_c$, we need to take $d=5$ or $d=6$. For
$d=5$, only the $+$ branch of the solution is allowed, and we can check that
around the corresponding solution $r_c$, $\frac{\partial^2 P_{\mathrm{GB}}}
{{\partial r_+^2}}$ indeed changes signature, thus the solution is indeed a
critical point. For $d=6$,  the two branches of solutions (\ref{RC})
degenerate, and $\frac{\partial^2 P_{\mathrm{GB}}}
{{\partial r_+^2}}$ does not change its signature around the corresponding
$r_c$, so we conclude that there is no critical point in six dimensions.

\subsubsection*{(1) Critical point in five dimensions}

\begin{figure}[h!]
\begin{center}
\includegraphics[width=0.45\textwidth]{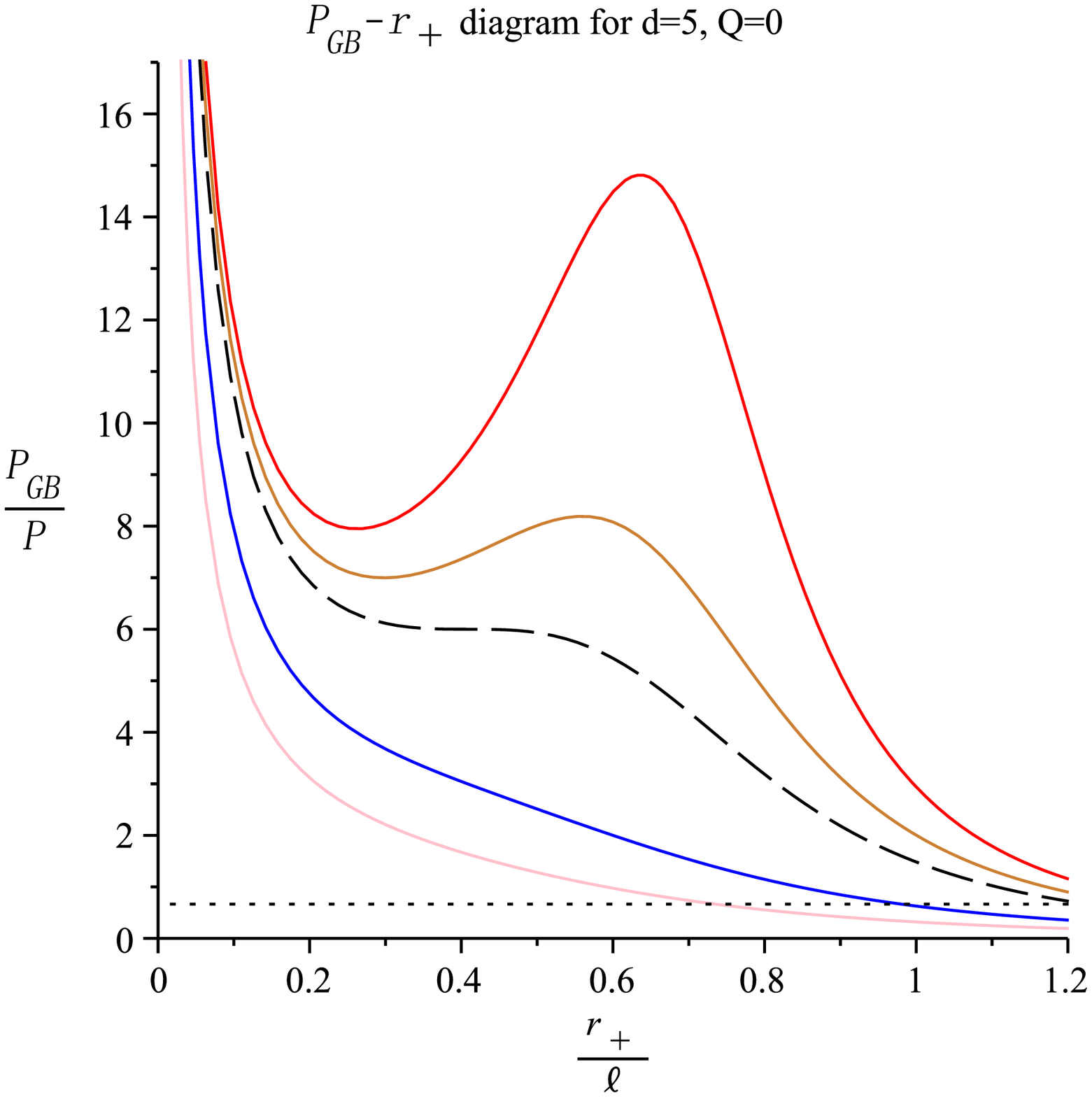}
\includegraphics[width=0.45\textwidth]{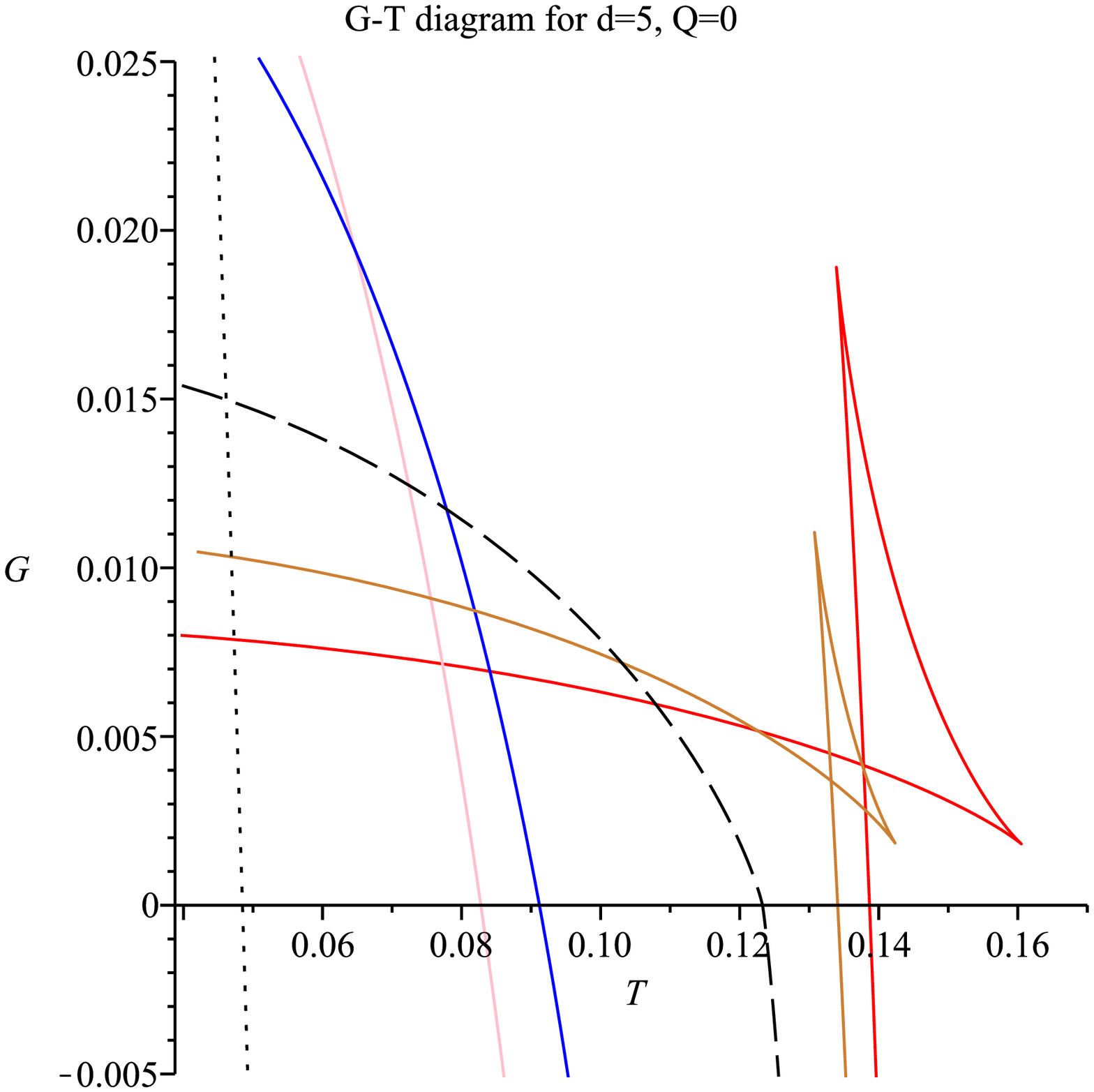}
\caption{The isotherm ($P_{\mathrm{GB}}-r_+$ plots at arbitrary 
constant $\ell$ on the left) and Gibbs free
energy at fixed $P_{\mathrm{GB}}$
($G-T$ plots at $\ell=\sqrt{10}$ on the right) for five
dimensional static neutral GB-AdS black holes with $k=1$. In both plots, the 
dotted line corresponds to that for the lower bound for $P_{\mathrm{GB}}$. The 
temperature of isotherms decrease from top to bottom. The lower
two isotherms contain no phase transition. The critical isotherm
$T=T_c$ is depicted in dashed line, while upper two isotherms correspond
to two-phase equilibrium states. The pressure $P_{\mathrm{GB}}$ on the right
plots increases from left to right, and the ``swallow tail"  behavior appears
only when $P_{\mathrm{GB}}>P_{\mathrm{GB}}^c$, which corresponds
to first order phase transition.}
\label{q0d5}
\end{center}
\end{figure}

In five dimensions, we have
\begin{align}
  R_c=\frac{1}{6},\quad  r_c=\frac{\sqrt{6}}{6}\ell,\quad
  P_{\mathrm{GB}}^c=\frac{9}{2\pi\ell^2},\quad  T_c=\frac{\sqrt{6}}{2\pi\ell},
  \label{crtv}
\end{align}
from which we can easily find
\begin{align}
\frac{P_{\mathrm{GB}}^c r_{c}}{T_c}=\frac{3}{2}.
\label{d5c1}
\end{align}
This relation is universal in the sense that it is independent of all
parameters. This result is very similar to the one in the Van der Waals system,
which has $\frac{P_cr_c}{T_c}=\frac{3}{8}$. This makes it more conceivable to
use the horizon radius $r_+$ instead of the thermodynamic volume
$V_{\mathrm{GB}}$ as an EOS parameter, since $V_{\mathrm{GB}}$ does not lead to
such a parameter independent relation. Since we are taking $P$ (i.e. $\ell$ via 
\eqref{pressure1}) as a constant parameter, it is better to work in units
of $P$ or $\ell$ and re-express the critical parameters in dimensionless form, 
\begin{align}
  \frac{r_c}{\ell}=\frac{\sqrt{6}}{6},\quad
  \frac{P_{\mathrm{GB}}^c}{P}=6.
  \label{crtv11}
\end{align}
In five dimensions, the previously mentioned lower bound for the dimensionless 
pressure is $\frac{P_{\mathrm{GB}}}{P}\geq\frac{2}{3}$, and the above critical 
value is well above this lower bound. So, the criticality can be always found 
in this case. We can have a clear look at this in the $P_{\mathrm{GB}}-r_+$ 
diagrams and $G-T$ diagrams which are presented in Fig.\ref{q0d5}.

It can be seen from Fig.\ref{q0d5} that, only for the isotherms with $T>T_c$, 
there exists a local minimum and maximum. Along
the segment of the isotherm between the these two extrema, we have
$\frac{\partial P}{\partial r_+}>0$, which implies that the black holes is in a
thermally unstable phase. For the $\frac{\partial P}{\partial r_+}<0$
regions, the black holes are thermally stable, corresponding
respectively to a small and a large black hole at the same temperature.
Physically speaking, the system at $T>T_c$ is in thermal equilibrium
between the stable small black hole and large black hole phases, because the
unstable, medium sized black hole phase cannot physically exist. The true
isotherm for such cases should be replaced by a steeply descending segment and
a slowly descending segment joined by a straight, horizontal (i.e. an isobar
$P_{\mathrm{GB}}=P^\ast=const.$) segment which can
be determined by Maxwell's equal area law.
When the critical temperature $T_c$ is reached, the shape of the isotherm will
undergo significant change. The two extrema merge into a single inflection
point, and we can no longer distinguish between the stable small and large
black holes. The isotherms with $T<T_c$ no longer contain any extrema and there
is only one branch with positive compression coefficient corresponding to
thermally stable black holes.

Turning to the Gibbs free energy plots, we see that each of the curves
corresponding to $P>P_{\mathrm{GB}}^c$ can extend to $T>T_c$ and contains a
``swallow tail" segment, which is a typical feature in first order phase
transitions.  From $P=P_c$ and downwards, the ``swallow tail" disappears, with
$P=P_c$ corresponding to the critical point.

\subsubsection*{(2) No critical point in six dimensions}

In six dimensions, we can obtain the following solution
\begin{align}
  R_c=\frac{1}{10},\quad  r_c=\frac{\sqrt{10}}{10}\ell,\quad
  P_{\mathrm{GB}}^c=\frac{5}{2\pi\ell^2},\quad  T_c=\frac{\sqrt{10}}{2\pi\ell}
  \label{RC2}
\end{align}
for the equations given in (\ref{critialpont}). However, it is easy to check
that $\frac{\partial^2 P_{\mathrm{GB}}}{{\partial r_+^2}}$ does not change its
signature around the above $r_c$, so, the above solution corresponds actually
to an extremum, rather than an inflection point on an isotherm. Consequently,
each isotherm contains one and only one extremum (which is a maximum). This
means that at each temperature, the value of $P_{\mathrm{GB}}$ has an upper
bound, black holes with $P_{\mathrm{GB}}$ bigger than the upper bound simply
could not exist. For smaller values of $P_{\mathrm{GB}}$, there are two
different black holes at each temperature: a small unstable black hole and a
large stable black hole. The small black hole phase cannot physically persist
because of its thermal instability, and there is no phase equilibrium in this
case.

\begin{figure}[h!]
\begin{center}
\includegraphics[width=0.45\textwidth]{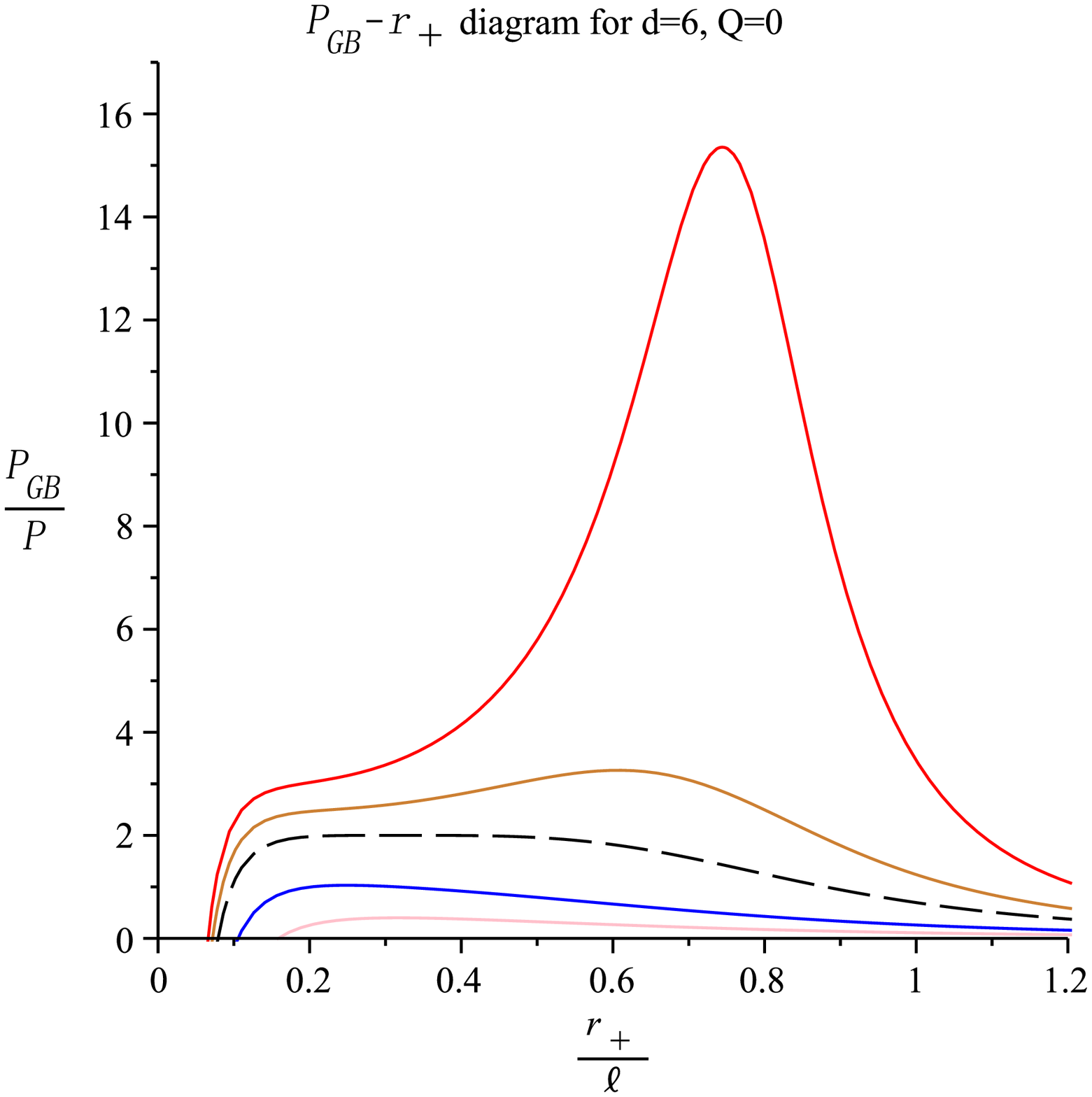}
\includegraphics[width=0.45\textwidth]{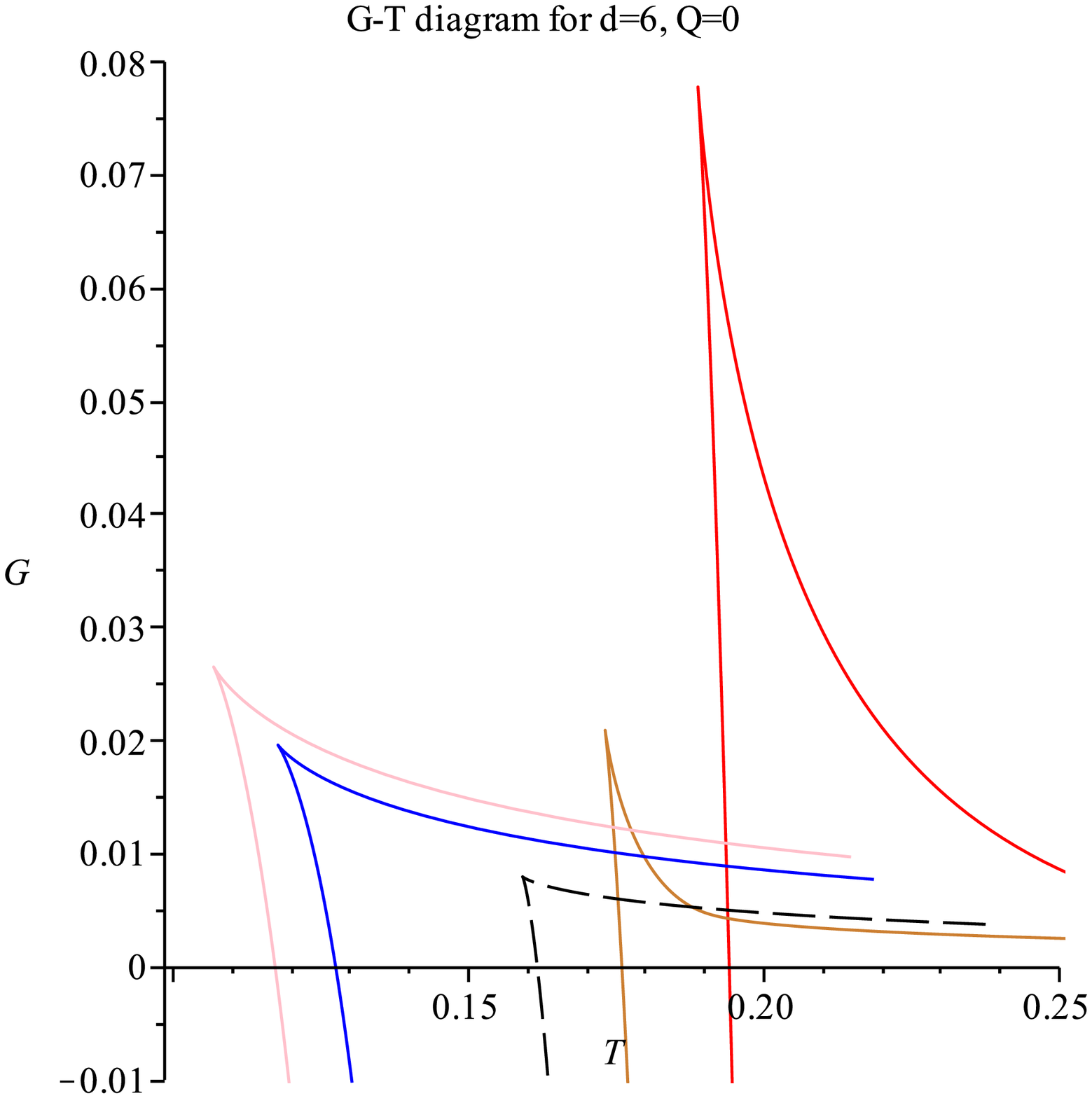}
\caption{The isotherm ($P_{\mathrm{GB}}-r_+$ plots at arbitrary constant $\ell$ 
on the left) and Gibbs free energy at fixed $P_{\mathrm{GB}}$  ($G-T$ plots at 
$\ell=\sqrt{10}$ on the right) for six
dimensional static neutral GB-AdS black holes with $k=1$. The
temperature of the isotherms decrease from top to bottom. There is no ``swallow
tail"  behavior in the $G-T$ diagram showing that there is no phase
equilibrium.}
\label{q0d6}
\end{center}
\end{figure}

In Fig.\ref{q0d6}, we present the $P_{\mathrm{GB}}-r_+$ and $G-T$ diagrams in
six dimensions. The dimensionless pressure and radius is used in the
$P_{\mathrm{GB}}-r_+$ plots. The dashed curves corresponding to the special 
values given in
(\ref{RC2}) plays no particular role as compared to other curves. It can be
seen on the $G-T$ plots that for each $P_{\mathrm{GB}}$, the $G(T)$ curve is
``$<$''-shaped, with the lower branch corresponding to the stable large black
hole phase.

\subsection{Critical exponents in five dimensions}

In this subsection, we will study the scaling behaviors of some physical
quantities near the criticality and compute the corresponding critical
exponents in five dimensions. The construction is analogous to the scaling
behavior of Van de Waals liquid-gas system, so we begin by reviewing the
scaling laws for Van de Waals system. Near the critical point, the critical
behavior of a Van der Waals liquid-gas system can be characterized by the
following critical exponents \cite{Reichl:1980}:
\begin{align}
  &C_v \sim  \left(\frac{|T-T_c|}{T_c}\right)^{-\alpha},  \\
  &\frac{v_g-v_l}{v_c}\sim  \left(-\frac{T-T_c}{T_c}\right)^{\beta},
  \label{beta} \\
  &\kappa_T\sim  \left(\frac{|T-T_c|}{T_c}\right)^{-\gamma}, \\
  &P-P_c\sim  (v-v_c)^{\delta},
  \label{cri-exp1}
\end{align}
where $C_v=T(\frac{\partial S}{\partial T})\big|_v$ is the the heat capacity at
constant volume, $\kappa_T=-v^{-1}(\frac{\partial v}{\partial P})\big|_T$ is
the isothermal compressibility, subscripts $g$ and $l$ stand for quantities in
the gaseous and liquid phases, respectively. The critical exponents take the
following values,
\begin{align}
  \alpha=0, \quad\beta=1/2,\quad \gamma=1, \quad\delta=3.
\end{align}

In our case, we need to replace the specific volume $v$ by the black hole
radius $r_+$ and study the corresponding scaling properties. The liquid and
gaseous phases for the Van der Waals system should also be replaced by the
small and large black hole phases, respectively. Since we shall be interested
in properties near the criticality,
we introduce the following dimensionless parameters which tend to zero near
the critical point,
\begin{align}
t=\frac {T} {T_c}-1,\quad \phi=\frac{r_+}{r_c}-1\quad
p=\frac {P_{\mathrm{GB}}}{P_{\mathrm{GB}}^c}-1.
\label{para}
\end{align}
Using these new parameters and inserting the critical values (\ref{crtv}) into
the EOS (\ref{PGB1}) at $d=5, k=+1$, we get the following {\em dimensionless}
EOS,
\begin{align}
p+1={\frac {t+1}{ \left( \phi+1 \right)  \left( -3\,\phi\,t-\phi-3\,t+1+
{\phi}^{2} \right) }}. \label{dless}
\end{align}
The Taylor series expansion for (\ref{dless}) at the critical point gives
\begin{align}
p&=4\,t+6\,t\phi- {\phi}^{3}+O \left(t\phi^2, {\phi}^{4} \right) ,
\label{pseries}
\end{align}
where we have neglected terms of order $\phi^4$ and $t\phi^2$ or higher as did
in  \cite{Kubiznak:2012wp}. It will be shown below that $t$ and $\phi^2$
are of the same order.

Using Maxwell's equal area law, we obtain the following equation
\begin{align}
0=\int_{\phi_l}^{\phi_s}\phi\frac{dp}{d\phi}d\phi\Rightarrow
- \frac3 2 (\phi_l^4-\phi_s^4)+6 t(\phi_l^2-\phi_s^2)=0,
\label{eqv1}
\end{align}
where the subscripts $s$ and $l$ stand for small and large stable black hole
phases, respectively. On the other hand, the physical phase equilibrium
condition (i.e. the isobar pressure condition) gives
\begin{align}
p|_{\phi_s}=p|_{\phi_l}\Rightarrow 6 t(\phi_l-\phi_s)-(\phi_l^3-\phi_s^3)=0.
\label{eqv2}
\end{align}
(\ref{eqv1}) and (\ref{eqv2}) together give a unique non-trivial solution
($\phi_s\neq\phi_l$)
\begin{align}
\phi_s=-\sqrt{6t},\quad \phi_l=\sqrt{6t}.
\label{phase}
\end{align}
We conclude that the coexistence of small and large black hole phases requires
$t>0$. In other words, only when $T>T_c$ can the two stable black hole phases
exist at the same pressure. This is different from that of a Van der Waals
liquid-gas system where $T<T_c$ is required for a phase equilibrium.
Eq. (\ref{phase}) can be rewritten as
\begin{align}
r_{+l} - r_{+s} \propto (T-T_c)^{1/2}. \label{rT}
\end{align}
This scaling behavior is in analogy to (\ref{beta}), which gives us
the critical exponent $\beta=\frac{1}{2}$.

The isothermal {\it radial} compressibility can be calculated as follows.
\begin{align*}
\kappa_T \equiv - \frac{1}{r_+} \frac{\partial r_+}{\partial P_{\mathrm{GB}}}
\bigg|_{r_+ =r_c}
\propto \left.-\frac1 {\frac{\partial p}{\partial \phi}} \right |_{\phi=0}
=-\frac1{6t}.
\end{align*}
This implies that
\begin{align}
\kappa_T\propto - (T-T_c)^{-1},
\label{gamma1}
\end{align}
which gives the critical exponent $\gamma=1$. In addition, it can be easily
seen that $p|_{t=0}=-\phi^3$, i.e.
\begin{align}
P_{\mathrm{GB}}-P_{\mathrm{GB}}^c \propto - (r_+ -r_c)^3,
\end{align}
which tells us $\delta=3$.

To evaluate the heat capacity near the critical point, we need to substitute
$d=5, k=+1$ into (\ref{IE}) and then calculate
\begin{align}
C_{r_+ \to r_c} \equiv \frac{\partial U}{\partial T}\bigg|_{r_+ \to r_c}
=-\frac{\sqrt{6}\,\ell}{48\pi^2T^2}.
\end{align}
This result is completely regular in $T$, showing that the critical exponent
$\alpha=0$. We see that though the EOS of our system is quite different from
that of the Van der Waals system, the resulting set of the critical exponents
are exactly the same as that for Van der Waals system.
It isn't a surprise that these critical exponents satisfy the following
thermodynamic scaling laws
\begin{align}
  &\alpha+2\beta+\gamma=2,\quad \alpha+\beta(1+\delta)=2, \nonumber \\
  &\gamma(1+\delta)=(2-\alpha)(\delta-1),\quad \gamma=\beta(\delta-1).
  \label{scaling}
\end{align}

\section{Criticality for static charged GB-AdS black holes}

\subsection{Critical behavior in five dimensions}

Having now understood the critical behavior of the static neutral GB-AdS black
holes associated with the new thermodynamic variables $P_{\mathrm{GB}}$
and $r_+$, we now turn
our attention to the static charged GB-AdS black holes. The analysis will be
basically parallel to the neutral cases, though the details are more
complicated due to the presence of extra parameters.

When $d = 5$, the numerator of the EOS (\ref{PGB2}) can be simplified.
Therefore we will discuss the two cases $d = 5$ and $d > 5$ separately. First
consider the case $d = 5$. The EOS (\ref{PGB2}) reduces to
\begin{align}
P_{\mathrm{GB}}=\frac{3 {kT\,r_+}}{6r_+^{2}
\left( {2\frac {r_+^2}{{\ell}^{2}}}+k-2\,T
\pi \,r_+ \right)-\frac{Q^2}{2r_+^{2}}}.
\label{PGB3}
\end{align}
This simplified EOS allows us to study the critical behavior analytically.
Inserting (\ref{PGB3}) into (\ref{critialpont}), one
finds that the critical horizon radius has to satisfy the following equation,
\begin{align}
  24R_c^3-4k R_c^2-\frac{5Q^2}{\ell^4}=0,\quad R_c\equiv \frac{r_c^2}{\ell^2}.
  \label{rcroot}
\end{align}
This equation has three analytical roots
\begin{align}
  &R_{c1}=\frac{1}{\ell^2}\left(\frac{1}{36}x_c^{1/3}+\frac{1}
  {9}\frac{k^2\ell^4}{x_c^{1/3}}+\frac{1}{18}k\ell^2\right),\label{rc1}\\
  &R_{c2}=\frac{1}{\ell^2}\left(-\frac{1}{2}\bigg[\frac{1}{36}x_c^{1/3}+
  \frac{1}{9}\frac{k^2\ell^4}{x_c^{1/3}}\bigg]+i\frac{\sqrt{3}}
  {2}\bigg[\frac{1}{36}x_c^{1/3}-\frac{1}{9}\frac{k^2\ell^4}{x_c^{1/3}}\bigg]+
  \frac{1}{18}k\ell^2\right),\nonumber\\
  &R_{c3}=\frac{1}{\ell^2}\left(-\frac{1}{2}\bigg[\frac{1}{36}x_c^{1/3}+
  \frac{1}{9}\frac{k^2\ell^4}{x_c^{1/3}}\bigg]-i\frac{\sqrt{3}}
  {2}\bigg[\frac{1}{36}x_c^{1/3}-\frac{1}{9}\frac{k^2\ell^4}{x_c^{1/3}}\bigg]+
  \frac{1}{18}k\ell^2\right),\nonumber
\end{align}
where
\begin{align}
  x_c=4860\,{Q}^{2}\ell^{2}+8k^3\,\ell^{6}+36\sqrt{15}\,\sqrt {1215\,
  {Q}^{2}+4k^3\ell^{4}}Q\ell^2.
\end{align}
Not all of these roots are real and positive, and we need to choose the
real positive root by some physical arguments.

Now assume that $R_c$ is given. Then the other critical quantities can be
evaluated with ease,
\begin{align}
  &T_c=\frac{3Q^2+4kr_c^4}{8r_c^5\pi}=\,{\frac {k+9\,R_c}{5
  \pi\ell\sqrt{R_c} }}, \label{Tfive}\\
  &P_{\mathrm{GB}}^c=P_{\mathrm{GB}}|_{r_+=r_c,T=T_c}={\frac {3}{20\pi}}\,
  {\frac { k+9\, R_c}
  {{\ell}^{2} k R_c\left( k-3\, R_c\right) }}, \label{Pfive}\\
  &\frac{P_{\mathrm{GB}}^cr_c}{T_c}=\,{\frac {3k}{4(k
  -3\, R_c)}}. \label{Rfive}
\end{align}
Unfortunately, the right hand side of (\ref{Rfive}) does not look so neat as in
(\ref{d5c1}), because it depends on the physical parameter $R_c$, or
alternatively, on $Q$ and $\ell$, thanks to the relation (\ref{rcroot}).
The critical temperature $T_c$ and pressure $P_{\mathrm{GB}}^c$ must be
positive. This leads to
\begin{align}
  &k+9\, R_c>0,\\
  &k\left( k-3\, R_c \right)>0.
\end{align}
For $k = 1$, $R_c$ must lie in the range
$0 < R_c < \frac{1}{3}$; For $k =-1$, $R_c$ must obey $R_c > \frac{1}{9}$.

On the other hand, the squared charge $Q^2$ can be taken as a function of
$R_c$ at the critical point due to (\ref{rcroot}),
\begin{align}
  Q^2\big(R_c\big)=\,{\frac {4R_c^2\ell^4 \left( 6R_c-k \right) }{5}}.
\end{align}
For $k = +1$, positivity of the squared charge gives a tighter bound for $R_c$,
i.e. $\frac{1}{6}<R_c<\frac{1}{3}$. For $k=-1$, the bound on $R_c$ does
not get tighter, however $Q^2$ is still bounded:
$Q > \frac{2\sqrt{3}}{27}\ell^2$. Summarizing the above discussions, we have
\begin{enumerate}
\item for $k=1$ (spherical horizon), the critical horizon radius
and charge need be in this region:
  \begin{align}
   \frac{1}{6}\leq R_c<\frac{1}{3},\quad 0\leq Q<\frac{2\sqrt{5}}
   {15}\ell^2;\label{region1}
  \end{align}
\item for $k=-1$ (hyperbolic horizon), the bounds are given as follows:
\begin{align}
   R_c>\frac{1}{9},\quad Q>\frac{2\sqrt{3}}{27}\ell^2.\label{region2}
  \end{align}
\end{enumerate}
In both cases we have $1215 Q^2 + 4k^3\ell^4 > 0$. This in turn implies that
only the root (\ref{rc1}) of (\ref{rcroot}) is real positive, and we always
have only one critical point in five dimensions.

\begin{figure}[h!]
\begin{center}
\includegraphics[width=0.45\textwidth]{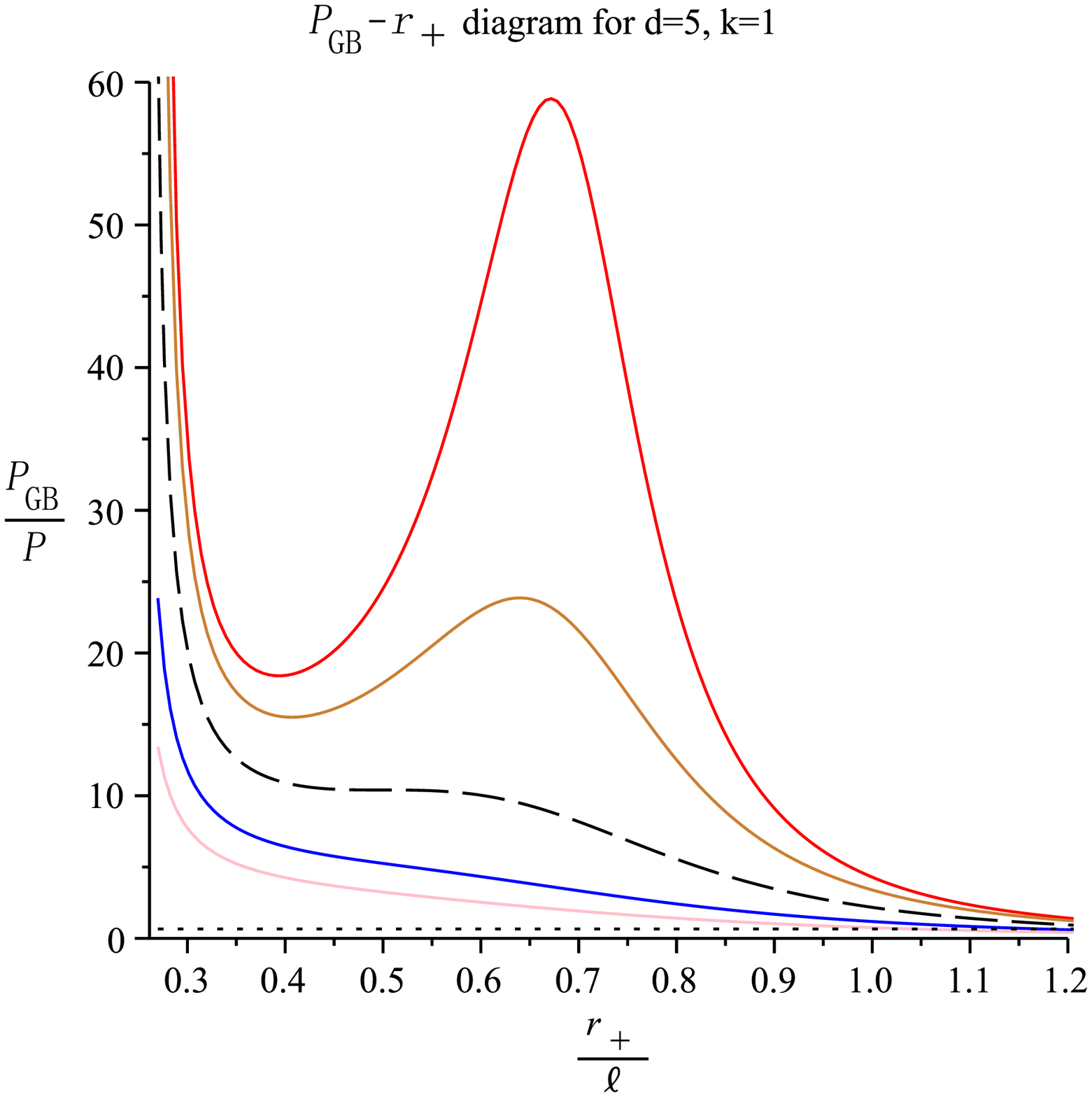}
\includegraphics[width=0.45\textwidth]{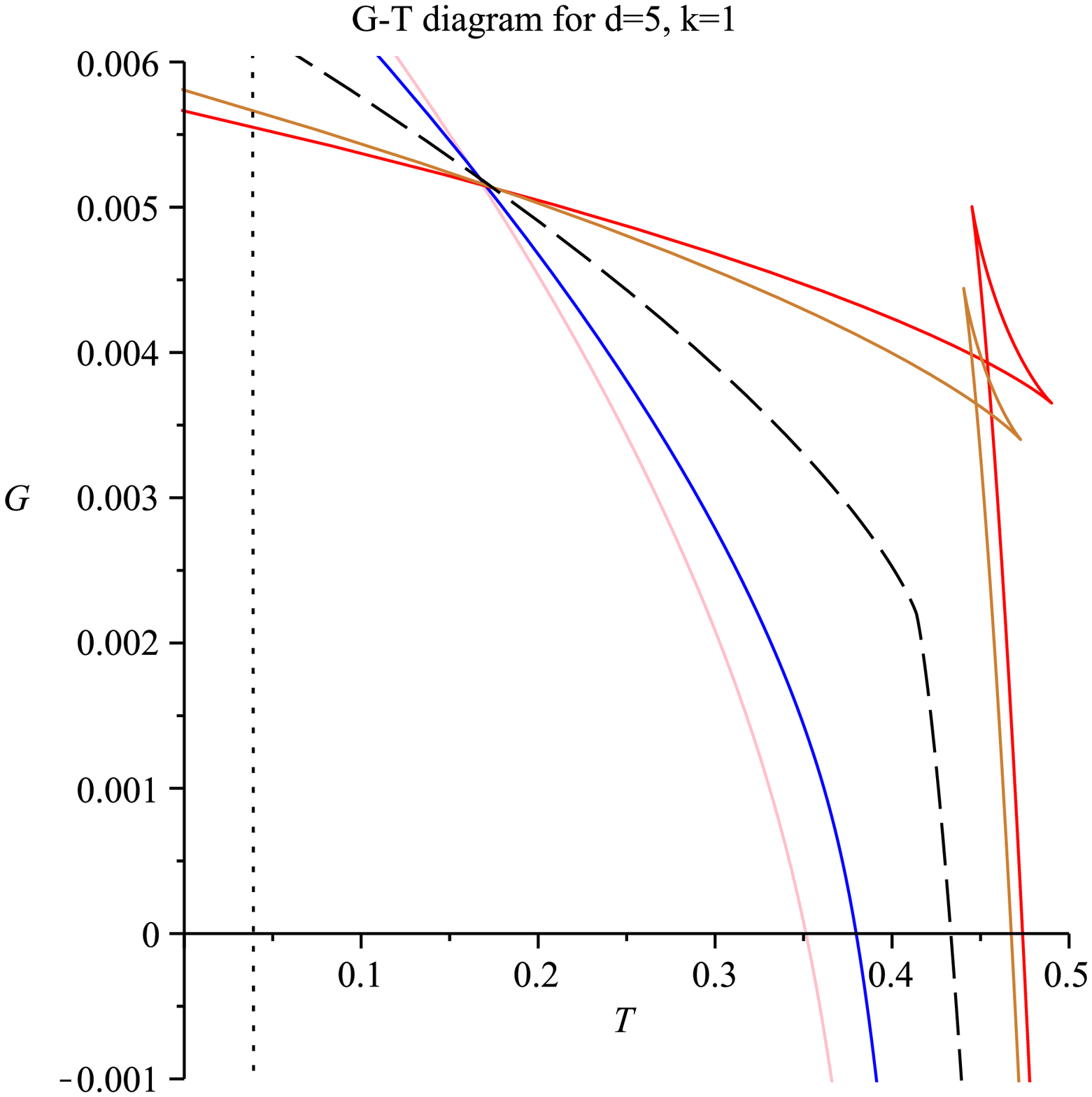}
\caption{The $P_{\mathrm{GB}}-r_+$ (at arbitrary constant $\ell$ on the left) 
and $G-T$ (at $\ell=8$ on the right) diagrams of
five dimensional static charged GB-AdS black holes at
$k=1$ and $Q=\frac{\sqrt{10}}{20}\ell^2$. On the left plots, the
temperature of isotherms decrease from top to bottom, with the dashed line
being the isotherm at the critical temperature and the dotted line 
corresponding to that for lower bound for $P_{\mathrm{GB}}$. On the right 
plots, the ``swallow tail"  behavior appears only when $P>P_c$, which 
correspond to the phase transition.}
\label{d5k1}
\end{center}
\end{figure}

In Fig.\ref{d5k1}, we depict the $P_{\mathrm{GB}}-r_+$ (at arbitrary constant 
$\ell$) and $G-T$ curves (at $\ell=8$)
for the special choice $k=+1$, and $Q=\frac{\sqrt{10}}{20}\ell^2$.
One finds that the ``swallow tail"
behavior appears only when $P>P_c,\, T>T_c$, which corresponds to phase
transition. Such phase transition is first order for $T>T_c$, while it becomes
second order at $T=T_c$ just as in the case of Van der Waals
system. In this case, the phase transition can always be found, as the pressure 
at the phase transition point is much bigger than the lower bound for the 
pressure as shown in Fig.\ref{d5k1}. The critical behavior is very similar to 
the case of neutral black holes. We can even show that the critical exponents 
are kept unchanged in the presence of electric charge, as will be shown for 
generic values of $Q$.

The dimensionless EOS in terms of the parameters defined in (\ref{para}) is
very complicated in the presence of electric charge $Q$. However, its Taylor
series expansion is simple enough and can be expressed as follows:
\begin{align}
p&=a_{10}t+a_{11}t\phi+a_{03}\phi^3 + O(t\phi^2,\phi^4),
\end{align}
where
\begin{align}
  a_{10}=\frac {4(2k+3R_c)}{5(k-3R_c)},\quad a_{11}=\frac {6(k+9R_c)}
{5(k-3R_c)},\quad a_{03}=\frac {2(k-9R_c)}{k-3R_c}.
\end{align}

To calculate the critical exponents, we shall follow the same
procedure that had led to (\ref{phase}). It follows that
\begin{align}
\phi_s=-\sqrt{-\frac{a_{11}}{a_{03}}t}
=-\sqrt{-\frac {3(k+9R_c)}{5(k-9R_c)}t},
\quad
\phi_l=\sqrt{-\frac{a_{11}}{a_{03}}t}=\sqrt{-\frac {3(k+9R_c)}{5(k-9R_c)}t}.
\label{phase-2}
\end{align}
For all allowed values of $R_c$ and $Q$ satisfying the bound (\ref{region1}) or
(\ref{region2}), we have always $\frac{a_{11}}{a_{03}}=
\frac {3(k+9R_c)}{5(k-9R_c)}<0$,
therefore the coexistence of the small and large black hole phases
requires $t>0$, i.e. the phase transition appears only at
temperature higher than $T_c$. The relation (\ref{rT}) still holds and the
critical exponent $\beta=\frac{1}{2}$.

The isothermal compressibility can be calculated easily, giving rise to
$\kappa_T\propto-\frac1{a_{11}t}$,
which indicates that the critical exponent $\gamma=1$. In addition, it can be
shown that $p|_{t=0}=a_{03}\phi^3$, which gives the critical exponent
$\delta=3$. Finally, $C_{r_+\to r_c}=-\frac{kr_c(3r_c^3+2k\ell^2)}
{20\pi^2\ell^2T^2}$  has no singular behavior at the critical points, and so
$\alpha=0$. This completes the proof that the critical exponents are kept
unchanged even in the presence of electric charge in five dimensions.

\subsection{Critical behavior in $d>5$ dimensions}

The critical behavior in $d>5$ dimensions is more difficult to analyze,
not only because of more complicated EOS, but also because that the conditions
(\ref{critialpont}) for the critical points become a much more complicated
set of algebraic equations, making it harder to solve analytically. To
understand such
criticality, we shall proceed in two different ways: For generic spacetime
dimension $d>5$, we shall take some particular choice for the
electric charge $Q$, which allows us to obtain the critical point parameters
analytically. On the other hand, in six dimensions, we shall study the
isotherms and $G-T$ curves numerically, taking $k=+1$ and $Q=0.01\ell^3$ at 
arbitrary constant $\ell$. This allows
us to understand the phase structure in this particular dimension.

\subsubsection{Analytical critical point in generic dimensions $d>5$}

In generic dimensions $d>5$, the critical point conditions (\ref{critialpont})
become a set of very complicated algebraic equations in $T_c$ and $r_c$.
To gain some insights into the solutions, we take the following particular
choice for the electric charge $Q$,
\begin{align}
  Q=\sqrt{\varrho}\,r_c^{d-3}, \label{Qrho0}
\end{align}
or, in terms of the dimensionless parameter $R_c = \frac{r_c^2}{\ell^2}$,
\begin{align}
Q=\sqrt{\varrho}\,R_c^{(d-3)/2}\,\ell^{d-3},
  \label{Qrho}
\end{align}
where $\varrho$ is a dimensionless parameter
because $Q$ has the dimension $[\mathrm{length}]^{d-3}$.

With the aid of the computer algebra system {\em Maple}, we can eliminate
$T_c$ from the pair of equations that follow from (\ref{critialpont}),
which yield a single algebraic equation for $R_c$,
\begin{align}
  &72 \left( d-1 \right) ^{2} \left( d-2 \right) ^{2}\,{R_c}^{2}\nonumber\\
  &\quad - \bigg( 6
\, \left( d-1 \right)  \left( d-2 \right)  \left( 5\,d-14 \right)
 \left( 2\,d-7 \right) \varrho +24\,k \left( d-1 \right)  \left( 2\,d-
9 \right)  \left( d-2 \right) ^{2} \bigg) R_c\nonumber\\
&\quad + \left( 2\,d -5\right)
 \left( d-4 \right)  \left( 2\,d-7 \right) ^{2}{\varrho }^{2}
 +k\left( d-2 \right)  \left( 14\,{d}^{3}-177\,{d}^{2}+697\,d-858
 \right) \varrho \nonumber\\
 &\quad +2\,{k}^{2} \left( d-3 \right)  \left( 7\,d-39
 \right)  \left( d-2 \right) ^{2}=0.
\label{eqcri}
\end{align}
The particular choice (\ref{Qrho}) for the electric charge makes the resulting
equation (\ref{eqcri}) of second order in
$R_c$, which is exactly solvable. The corresponding $T_c$ given by
\begin{align}
T_c &= \mathcal{N}^{-1}k \,(d-5)\left( \varrho \,d+2\,kd-4\,\varrho -6\,k
\right)   \nonumber\\
& \quad \times \left( k{d}^{2}-4\,\varrho \,{d}^{2}+18\,R_{{c}}{d}^{2}+13
\,kd+24\,\varrho \,d-54\,R_{{c}}d-30\,k-41\,\varrho +36\,R_{{c}}
 \right), \label{Tcd}
\end{align}
where
\begin{align}
\mathcal{N}&= \left( 52\,{k}^{2}{d}^{3}+10\,k\varrho \,{d}^{3}
-8\,{\varrho }^{2}{d}
^{3}-356\,{k}^{2}{d}^{2}-51\,k\varrho \,{d}^{2}+44\,{\varrho }^{2}{d}^
{2}+576\,{k}^{2}d+24\,kR_{{c}}{d}^{3}\right.\nonumber\\
&\quad -360\,kR_{{c}}{d}^{2}+912\,kR_{{c
}}d+23\,k\varrho \,d+12\,\varrho \,{d}^{3}R_{{c}}-30\,\varrho \,{d}^{2
}R_{{c}}+6\,\varrho \,dR_{{c}}-46\,{\varrho }^{2}d
\nonumber\\
&\quad \left.-144\,{k}^{2}+198\,k
\varrho -576\,kR_{{c}}-35\,{\varrho }^{2}+12\,\varrho \,R_{{c}}
 \right)\pi \,\ell \sqrt {R_{{c}}}.
\end{align}
For $d>5$, there can be up to two different critical points of
critical radius $r_c =\ell \sqrt{R_c}$ for a given $\varrho$ with
appropriate value. However, please bear in mind that these two critical
points do not correspond to the same electric charge, because the charge
$Q$ and $R_c$ are related via (\ref{Qrho}). Moreover,
in the expression (\ref{Tcd}) for $T_c$,
there is a factor $(d-5)$, which implies that both solutions to (\ref{eqcri})
correspond to $T_c=0$ when $d=5$. In fact, the five dimensional
critical point with $T_c \neq 0$ described in the previous subsection does
not correspond to any of the solutions of (\ref{eqcri}) with $d=5$.

The explicit solutions to (\ref{eqcri}) read
\begin{align}
R_{c\pm} &=\frac{1}{6(d-1)}\left[(2d-9)k+\frac{1}{4}(2d-7)(5d-14)\varrho
\pm \sqrt{3\Delta}\right],\\
\Delta &= 3\, \left( 2\,d-7 \right) ^{2}{\varrho }^{2}+8\,k \left( 2\,d-1
 \right)  \left( d-4 \right) \varrho -16\,{k}^{2} \left( d-2 \right)
 \left( d-6 \right).
\end{align}
We can check that for $d>5$,  $\frac{\partial^2 P_{\mathrm{GB}}}
{{\partial r_+^2}}$ changes signature around both solutions $r_{c\pm}$, thus
the solutions indeed correspond to critical points.
Using this solution, all other critical parameters can be evaluated with ease,
though the concrete results are too complicated to be reproduced here. Please
note that the real positivity of $R_{c\pm}$ and the corresponding $T_{c\pm}$
and $P_{\mathrm{GB}}^{c\pm}$ naturally imposes some bounds on $\varrho$, so
the two critical points will appear only for electric charges within some
specific region, just like in the previously studied
five dimensional case.

\subsubsection{Numerical results in $d=6$ dimensions}

\begin{figure}[h!]
\begin{center}
\includegraphics[width=0.45\textwidth]{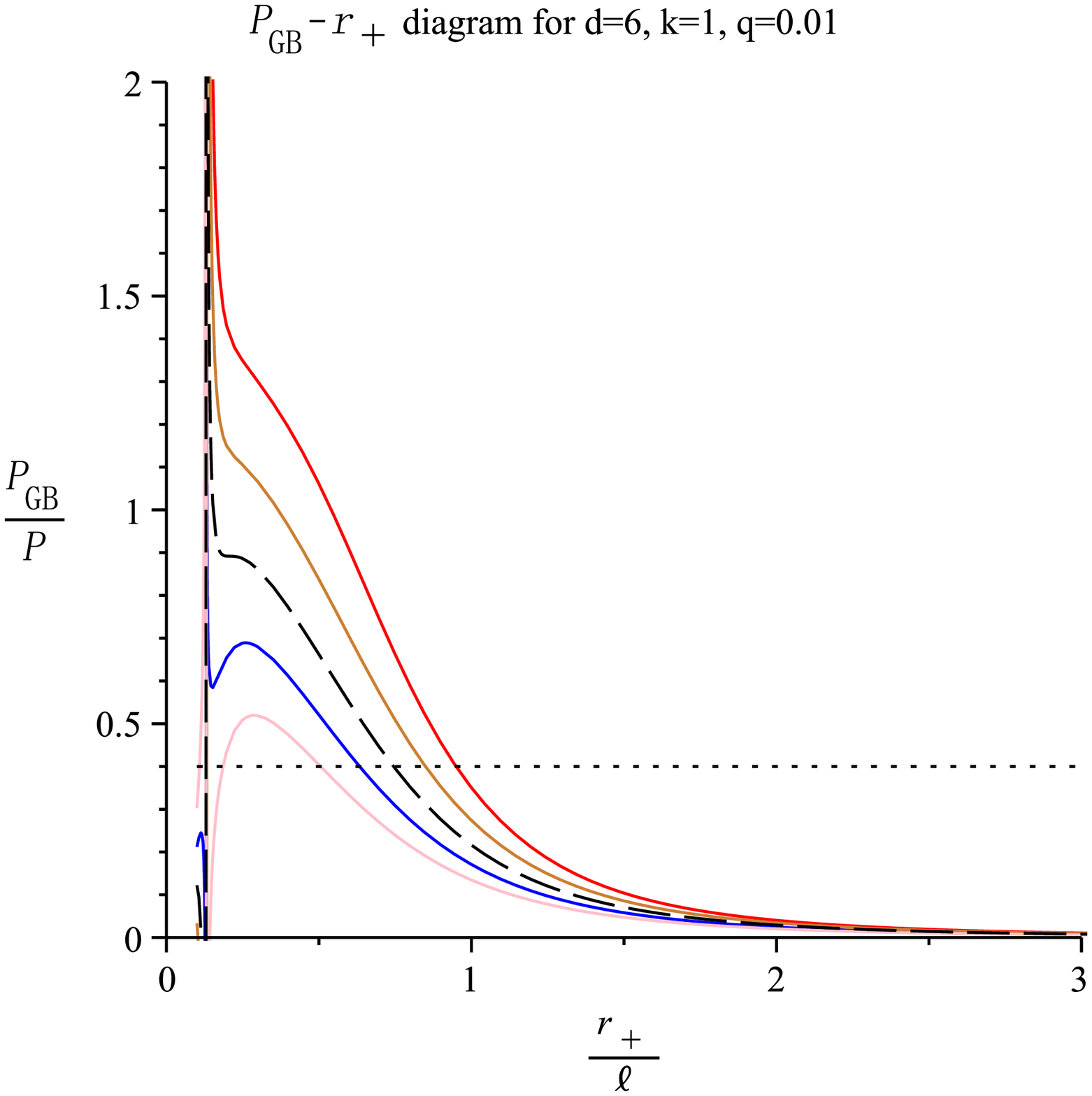}
\includegraphics[width=0.45\textwidth]{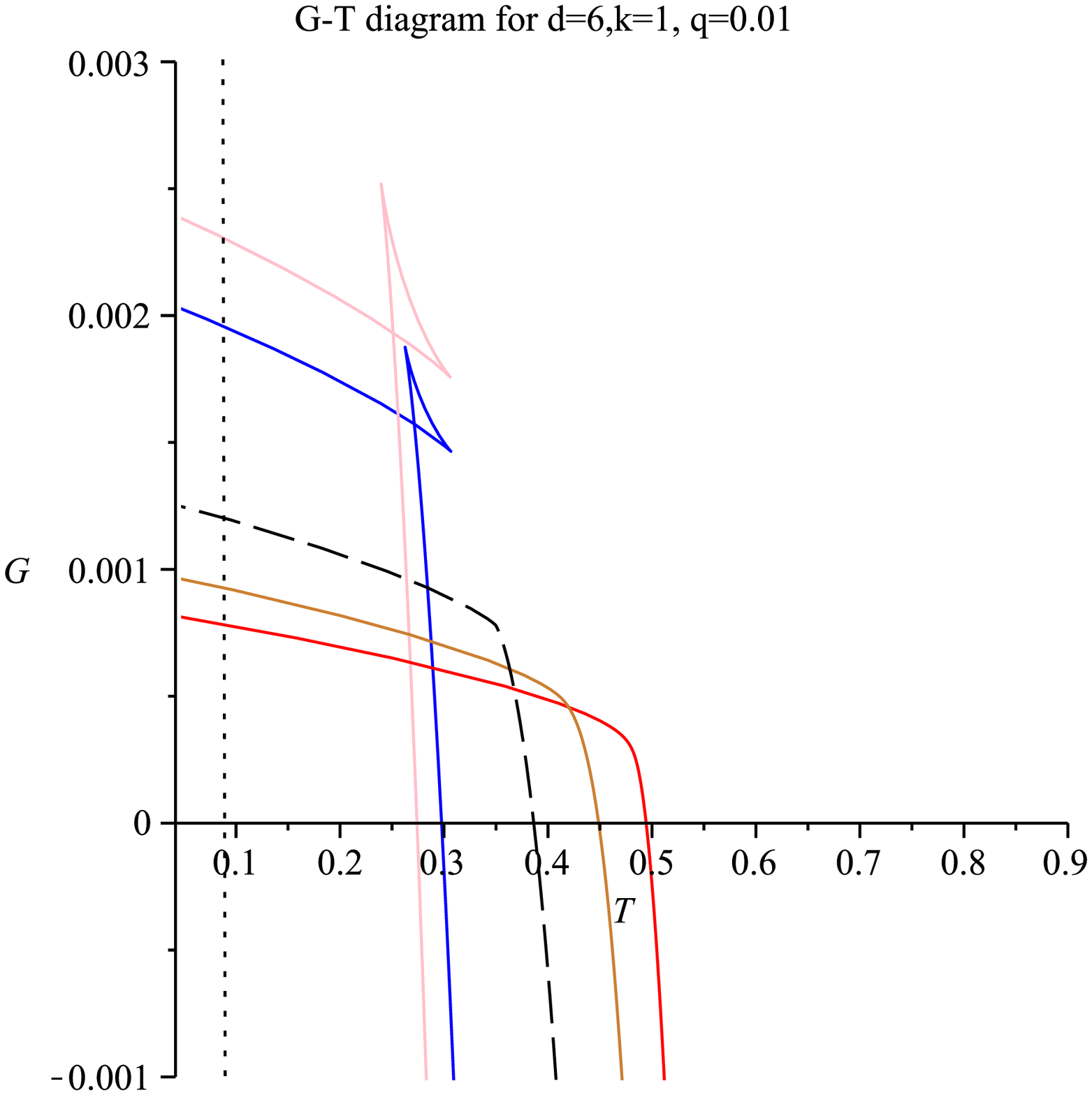}
\caption{The $P_{\mathrm{GB}}-r_+$ (at arbitrary constant $\ell$ on the left) 
and $G-T$ (at $\ell=1$ on the right) diagrams for six
dimensional static charged GB-AdS black holes with $q=0.01, k=1$.
The isotherms are all near the lower critical temperature $T=T_s$
and the $G-T$ curves are all near the corresponding critical pressure
$P_{\mathrm{GB}}=P_{\mathrm{GB}}^s$. The dashed lines correspond to the
critical curves and the dotted lines correspond to the lower bound for the 
pressure.}
\label{d6k1q001one}
\end{center}
\end{figure}

\begin{figure}[h!]
\begin{center}
\includegraphics[width=0.45\textwidth]{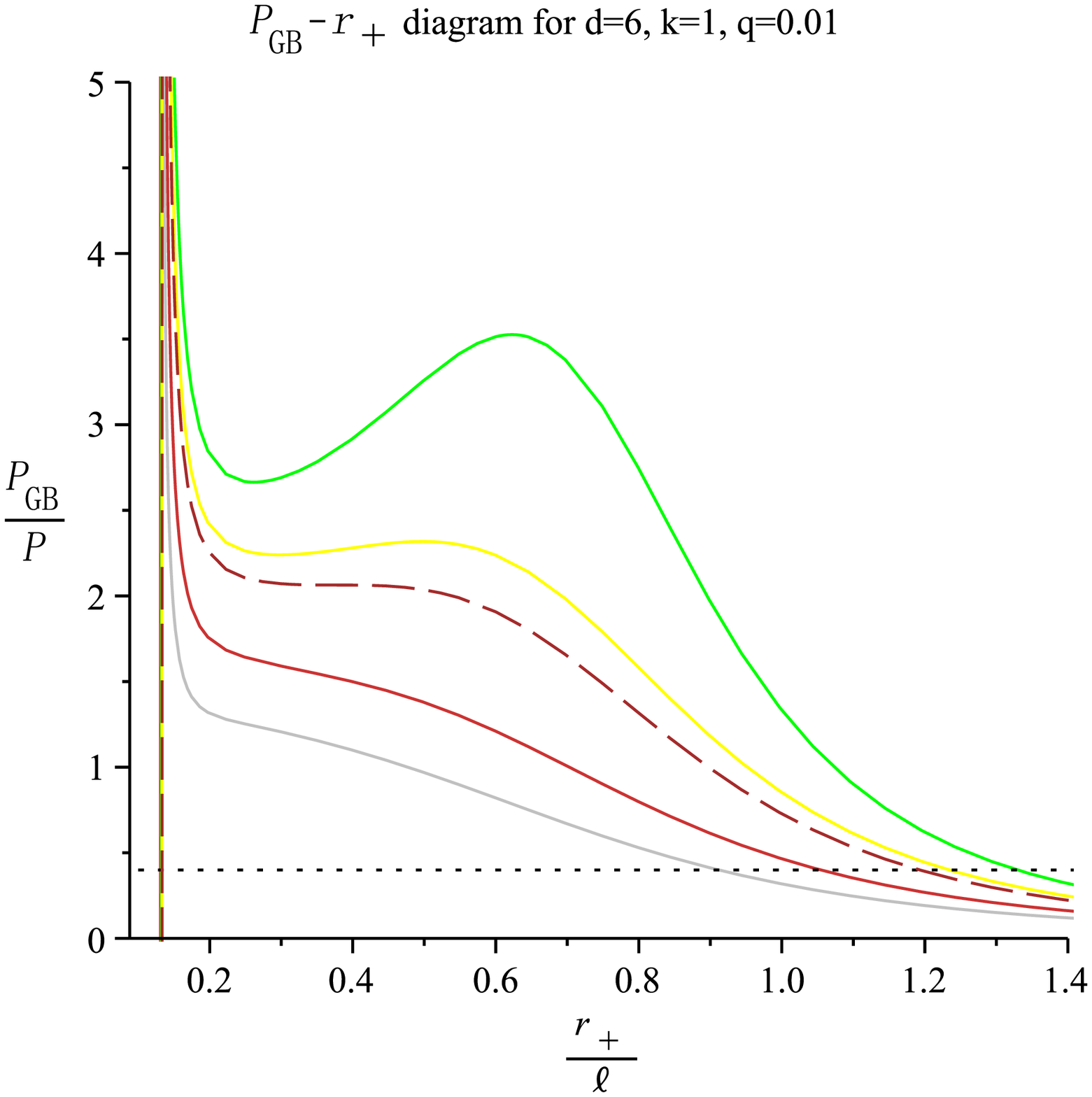}
\includegraphics[width=0.45\textwidth]{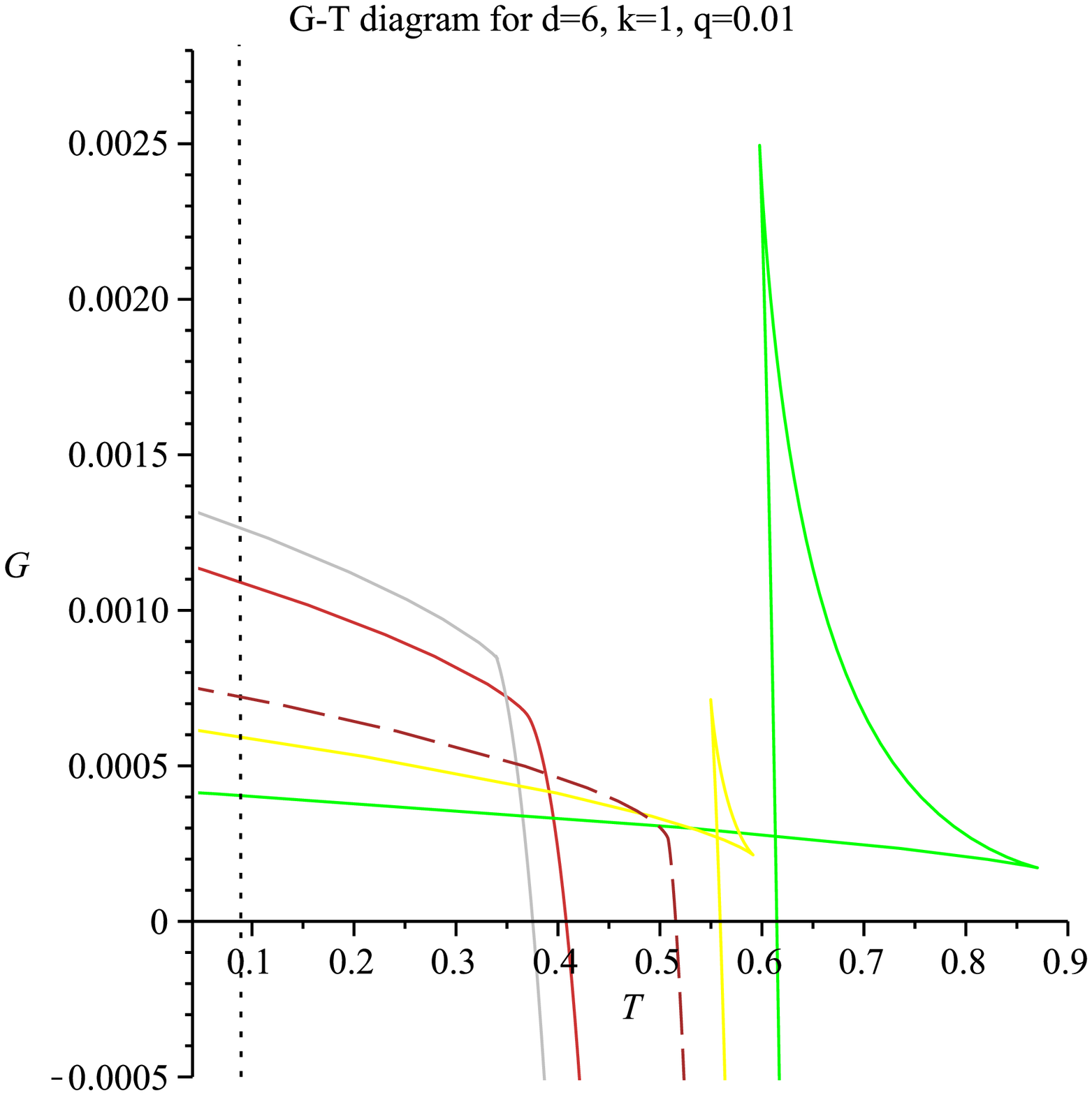}
\caption{The $P_{\mathrm{GB}}-r_+$ (at arbitrary constant $\ell$ on the left) 
and $G-T$ (at $\ell=1$ on the right) diagrams for six
dimensional static charged GB-AdS black holes with $q=0.01, k=1$.
The isotherms are all near the higher critical temperature $T=T_l$
and the $G-T$ curves are all near the corresponding critical pressure
$P_{\mathrm{GB}}=P_{\mathrm{GB}}^l$. The dashed lines correspond to the
critical curves and the dotted lines correspond to the lower bound for the 
pressure.}
\label{d6k1q001two}
\end{center}
\end{figure}

\begin{figure}[h!]
\begin{center}
\includegraphics[width=0.45\textwidth]{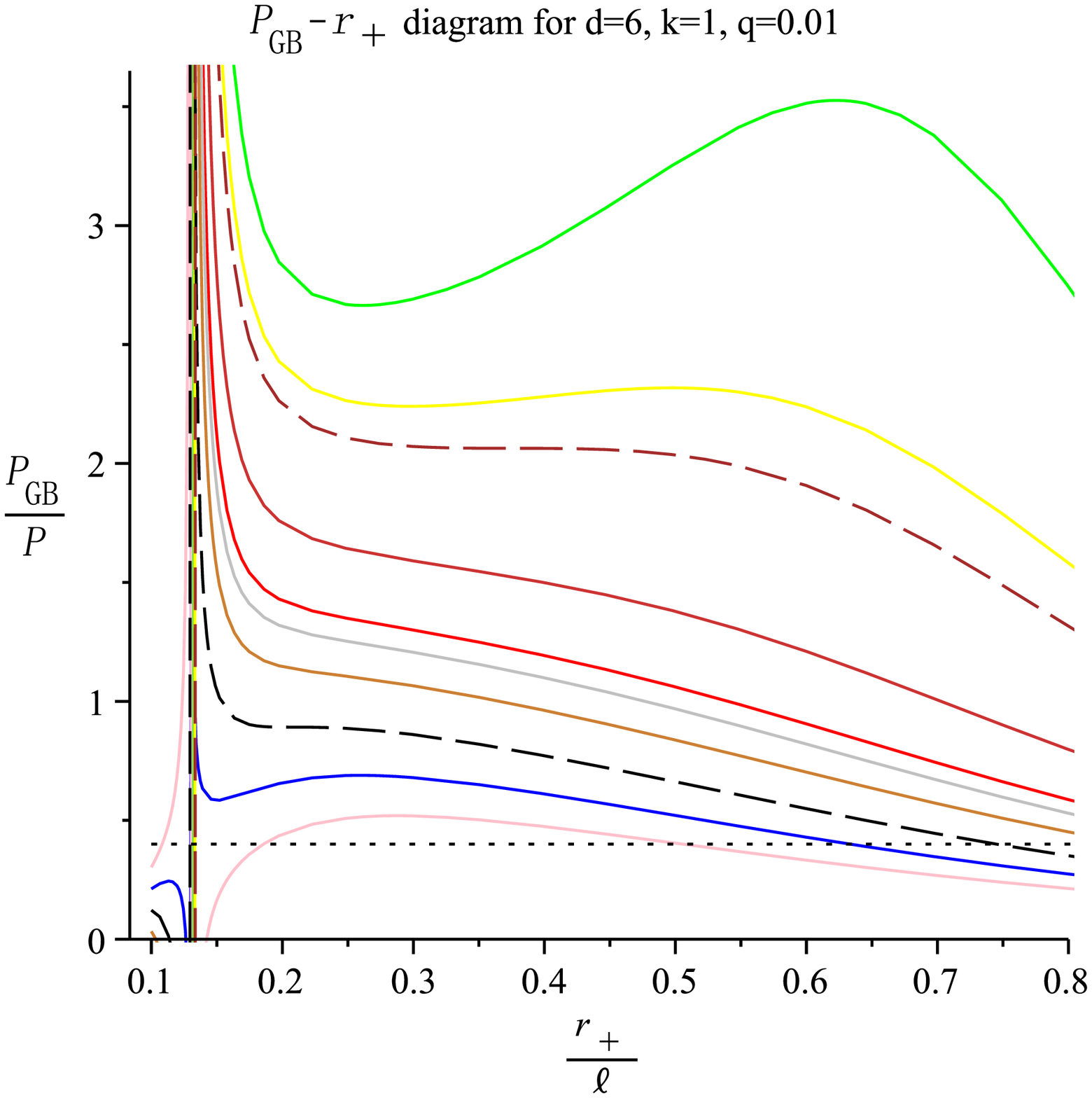}
\includegraphics[width=0.45\textwidth]{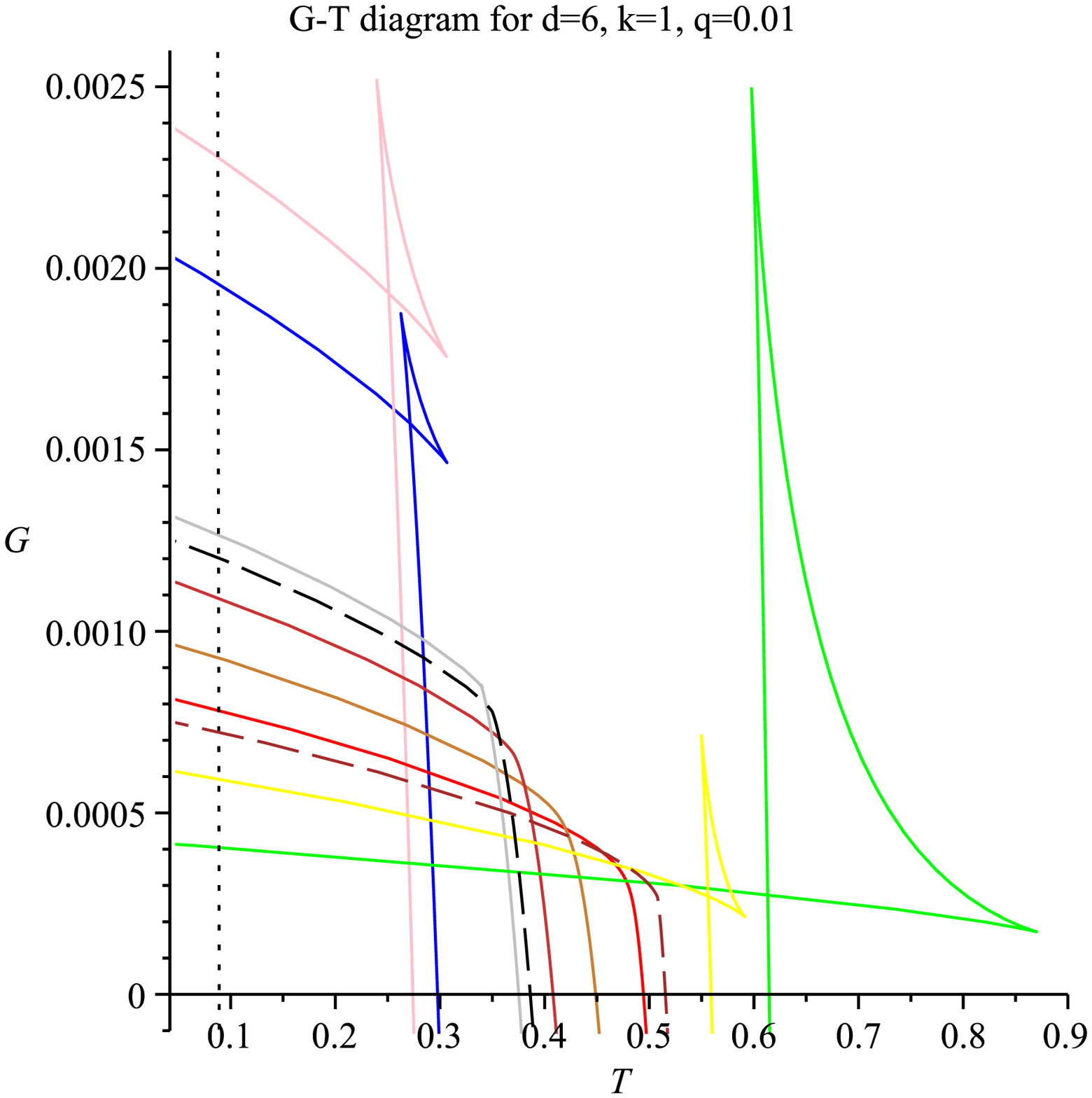}
\caption{The combination of Fig.\ref{d6k1q001one} and Fig.\ref{d6k1q001two}.
Note that the two critical isotherms on the left plots never cross each other,
however the corresponding $G-T$ curves do cross each other.}
\label{total}
\end{center}
\end{figure}

To analyze the critical points at fixed charge $Q$ rather than fixed $\varrho$,
we now work numerically in the case $d=6$. Assuming that the AdS radius $\ell$
is fixed and writing
\begin{align}
  Q=q\ell^3, \quad r_c=\sigma_c\ell,
\end{align}
where both $q$ and $\sigma_c$ are dimensionless, the critical point conditions
can be reduced into the following equation for the critical radius
parameter $\sigma_c$,
\begin{align}
  175\,{q}^{4}-48\sigma_c^6{q}^{2}
  -4800{q}^{2}\sigma_c^8+144\sigma_c^{12}-2880\sigma_c^{14}
  +14400\sigma_c^{16}=0.
\end{align}
Setting $q=0.01$, the critical radius parameter can be worked out numerically,
giving rise to two different critical radii: a small critical radius
$\sigma_s\approx 0.208$ with the corresponding critical temperature
$T_s \approx \frac{0.350}{\ell}$ and critical thermodynamic pressure
$P_{\mathrm{GB}}^s \approx \frac{0.355}{\ell^2}$ 
(i.e. $\frac{P_{\mathrm{GB}}^s}{P} \approx 0.892$) and a large critical radius
$\sigma_l \approx 0.377$ with the corresponding temperature $T_l \approx
\frac{0.508}{\ell}$ and pressure $P_{\mathrm{GB}}^l\approx\frac{0.821}
{\ell^2}$ (i.e. $\frac{P_{\mathrm{GB}}^s}{P} \approx 2.063$). The 
$P_{\mathrm{GB}}-r_+$ and $G-T$ diagrams around each of these
critical points are depicted in Fig.\ref{d6k1q001one} and
Fig.\ref{d6k1q001two}, respectively. One finds that the swallow tails appear
either at  $0.4P<P_{\mathrm{GB}}<P_{\mathrm{GB}}^s$ (i.e. $0.4<
\frac{P_{\mathrm{GB}}}{P}<0.892$), $T_{l}<T<T_s$ or at
$P_{\mathrm{GB}}>P_{\mathrm{GB}}^l$ (i.e. $\frac{P_{\mathrm{GB}}}{P}>2.063$), 
$T>T_l$ but not when $P_{\mathrm{GB}}^s<P_{\mathrm{GB}}<P_{\mathrm{GB}}^l$, 
(i.e. $0.892<\frac{P_{\mathrm{GB}}}{P}<2.063$), $T_s<T<T_l$. Here $P_{GB}=0.4P$ 
is precisely the lower bound for the pressure in six dimensions. When 
$T=T_{l}$, the pressure at the phase transition point happens to take value
at the lower bound.
The combination of Fig.\ref{d6k1q001one} and
Fig.\ref{d6k1q001two} is given in Fig.\ref{total}, which gives the
complete critical structure of the system at this particular value of
electric charge.

\section{Conclusions}

Taking the (inverse of) GB coupling $\alpha$ as a new thermodynamic variable
$P_{\mathrm{GB}}$, we revisited the thermodynamics for GB-AdS black holes and
studied the associated critical behavior at fixed electric charge and bare
cosmological constant. It is shown that for static neutral GB-AdS black holes,
the corresponding critical point exists only for black holes with spherical
topology (i.e. $k=+1$) in five dimensions, and the set of
critical exponents are identical to those of Van der Waals system. This is
quite similar to the $P-V$ criticality associated with the cosmological
constant at fixed GB coupling for the same black holes \cite{Cai:2013qga}.
However, there is a
crucial difference from the case of $P-V$ criticalities: in our case, the phase
transition occur only when the temperature is {\em higher} than the critical
temperature, while the phase transition in $P-V$ criticalities occur only when
the temperature is lower than the critical temperature. The situation for
static charged GB-AdS black holes is much more complicated, and it is
shown that there can only be one critical point in five dimensions (for either
$k=+1$ or $k=-1$) when the electric charge $Q$ obeys some appropriate
bound. The corresponding critical exponents are also identical to those for Van
der Waals system. In higher dimensions, it is shown that there can be two
critical points if the electric charge is taken to be proportional to the
$(d-3)$-th power of the critical radius ($Q\propto r_c^{d-3}$). Numerical study
also shows that in six dimensions, there can be two different critical points
at the same fixed electric charge, and the phase transitions can occur
when the temperature is either lower than the lower critical temperature or
higher than the higher critical temperature but not in between the two critical
temperatures. This situation is not seen in earlier studies on $P-V$
criticalities for the same theory. Therefore, our study indicates that there
are still much richer, unexpected structures in the thermodynamics of GB-AdS
black holes.

\begin{figure}[h!]
\begin{center}
\includegraphics[width=0.45\textwidth]{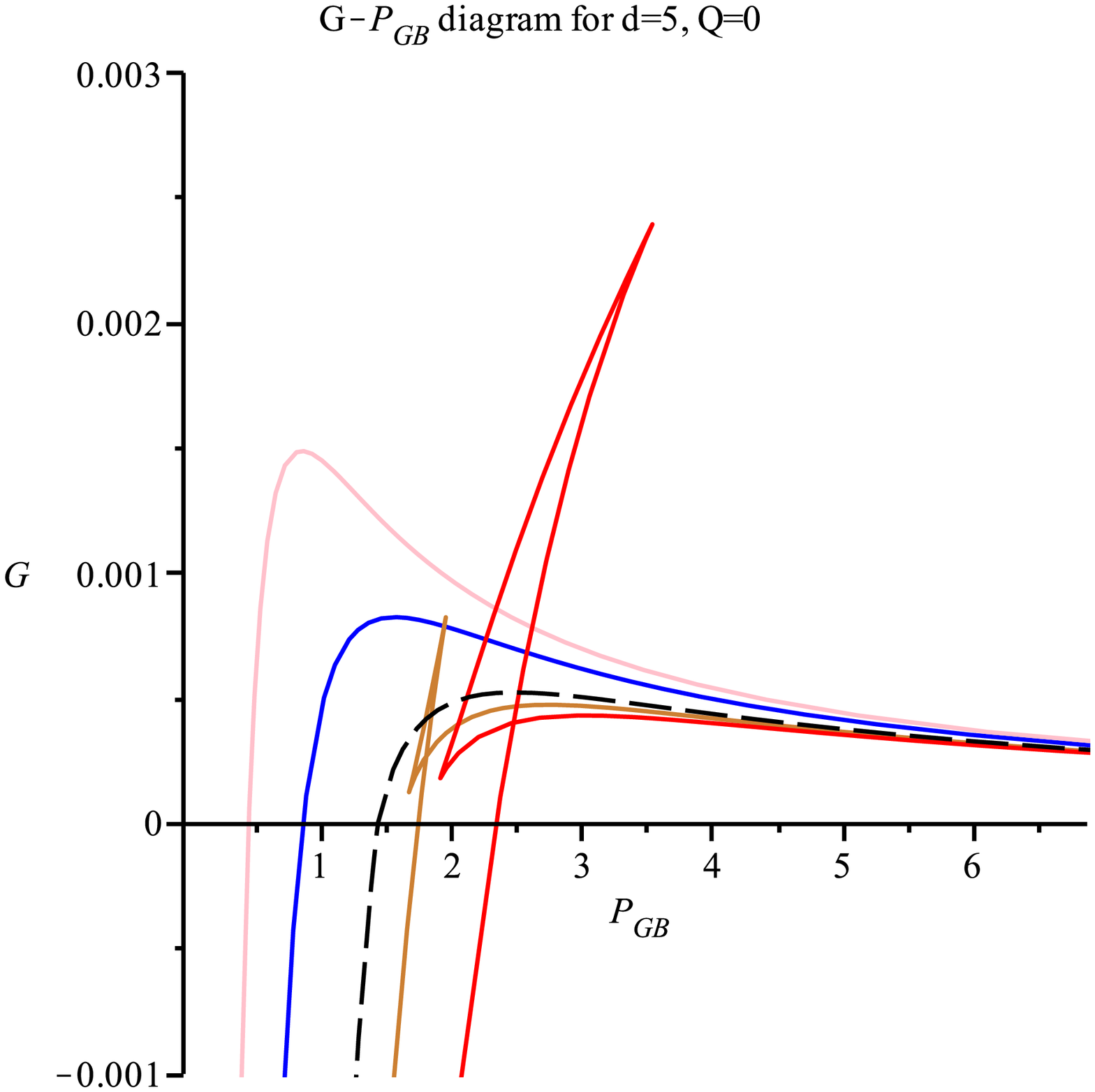}
\includegraphics[width=0.45\textwidth]{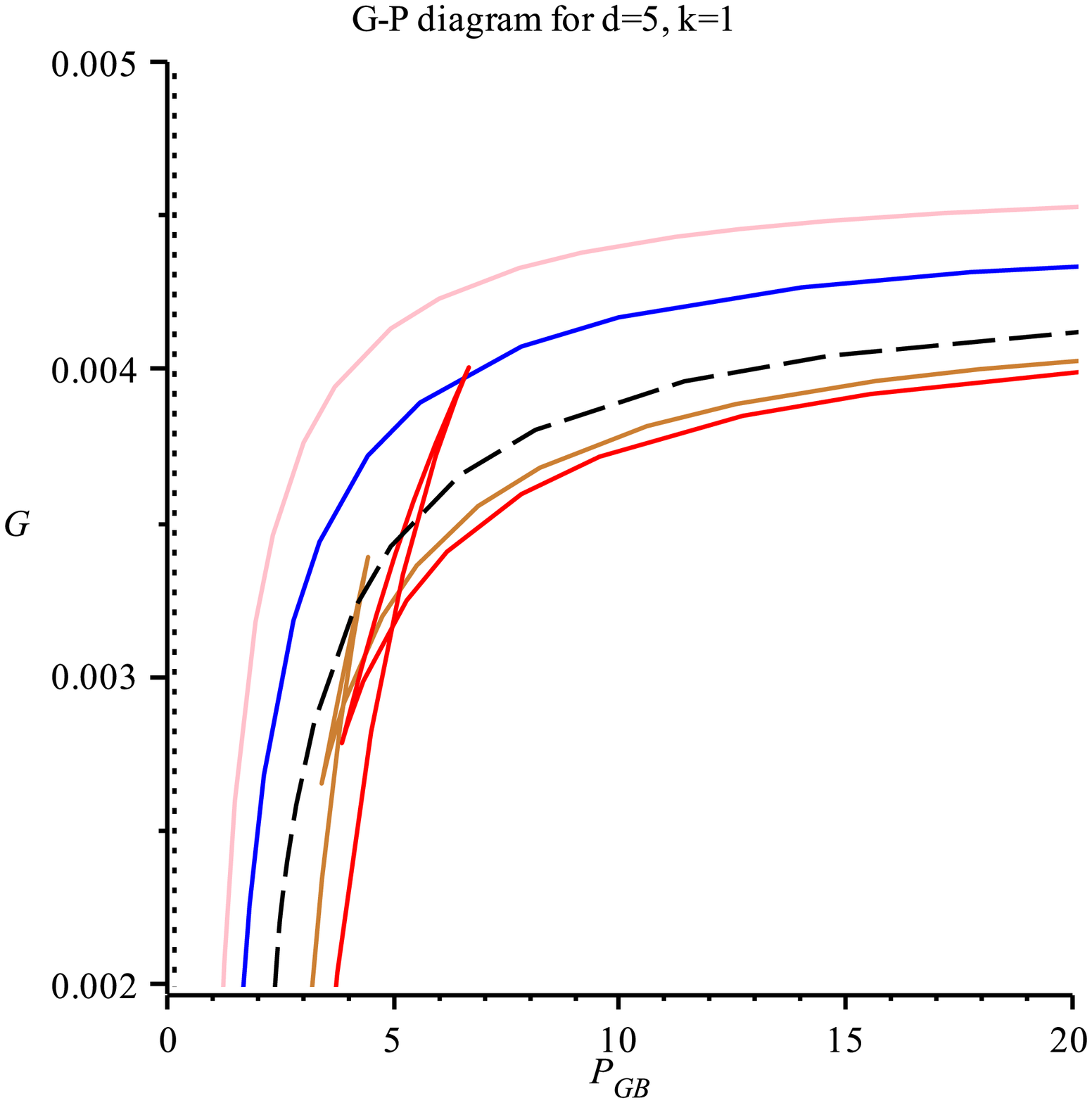}
\caption{The Gibbs free energy at fixed $T$ ($G-P_{\mathrm{GB}}$ plots) for 
five dimensional static neutral GB-AdS black holes (with $k=1,\ell=1$ on the 
left) and for five dimensional static charged GB-AdS black holes (with
$k=1, \ell=8$ and $Q=\frac{\sqrt{10}}{20}\ell^2$ on the right). The critical 
curves $T=T_c$ are depicted in dashed lines and the dotted lines correspond to 
that for the lower bound for the pressure. The temperature $T$ on the both
plots increases from left to right, and the ``swallow tail"  behavior appears
only when $T>T_c$, which corresponds
to first order phase transition.}
\label{d5GP}
\end{center}
\end{figure}

Before closing, let us make some comments on the relation of our results and 
the $\alpha\rightarrow 0$ limit, in which the theory reduces to Einstein 
gravity. When $\alpha$ is vanishing, $P_{\mathrm{GB}}$ is infinite and 
$V_{\mathrm{GB}}$ is identically zero, thus $P_{\mathrm{GB}}$ and 
$V_{\mathrm{GB}}$ lose their role as a pair of 
thermodynamic variables. This seems to invalidate our discussion at this 
particular limit. However, our results still stands because the criticalities 
described in this paper all appear at finite $P_{\mathrm{GB}}$ and so are 
invisible from the Einstein gravity limit.

It will be illuminating to look at the $G-P_{\mathrm{GB}}$ plots
at fixed $T$ which are presented in Fig.\ref{d5GP}. For each 
$G-P_{\mathrm{GB}}$ curve, the global minimum of the Gibbs free energy is
``$\Gamma$''-shaped, with the left steep branch corresponding to the stable 
large black hole phase (small $P_{\mathrm{GB}}$ and large $\alpha$) and the 
right gentle branch corresponding to the stable small black hole phase (large 
$P_{\mathrm{GB}}$ and small $\alpha$). One can also find the ``swallow tail"  
corresponding to first order phase transition which is always located near 
the small $P_{\mathrm{GB}}$ end. That to say, at fixed $T$, one can never find 
phase transition associated with the variable $P_{\mathrm{GB}}$ near the 
$\alpha\rightarrow 0$ limit (``Einstein gravity" black hole phase).
The work presented in this paper indicates that the $\alpha\rightarrow 0$ limit 
is a metastable phase of the theory if the GB coupling $\alpha$ is to be 
considered as a thermodynamic variable, and the large $\alpha$ (small 
$P_{\mathrm{GB}}$) phase is thermodynamically preferred. On the other hand, 
the existence of a lower bound for $P_{\mathrm{GB}}$ ensures that 
$P_{\mathrm{GB}}$ will not go to zero in a thermodynamic process, so 
one does not need to worry about the possibility that $\alpha$ runs to
infinity.

\providecommand{\href}[2]{#2}\begingroup
\footnotesize\itemsep=0pt
\providecommand{\eprint}[2][]{\href{http://arxiv.org/abs/#2}{arXiv:#2}}
\endgroup

\end{document}